\documentclass[twocolumn]{aastex631}

\newcommand{\sbi}{\texttt{swyft}}
\renewcommand{\arraystretch}{1.5}
\usepackage{booktabs}

\usepackage{amsmath}
\usepackage{makecell}
\usepackage{comment}
\usepackage{bm}

\shorttitle{SBI of radio MSPs in globular clusters}
\shortauthors{Berteaud et al.}
\graphicspath{{./}{figures/}}

\begin{document}

    \title{Simulation-based inference of radio millisecond pulsars in globular clusters}

\author[0000-0003-4962-145X]{Joanna Berteaud}
\affiliation{University of Maryland, 
Department of Astronomy, 
College Park, MD 20742, USA}
\affiliation{NASA Goddard Space Flight Center, 
Code 662, 
Greenbelt, MD 20771, USA}

\author[0000-0002-5135-2909]{Christopher Eckner}
\affiliation{Center for Astrophysics and Cosmology, University of Nova Gorica, Vipavska 11c, 5270 Ajdov\v{s}\v{c}ina,
Slovenia}
\affiliation{LAPTh, CNRS, F-74000 Annecy, France}
\affiliation{LAPP, CNRS, F-74000 Annecy, France}

\author[0000-0001-7722-6145]{Francesca Calore}
\affiliation{LAPTh, CNRS, F-74000 Annecy, France}

\author[0000-0003-0724-2742]{Ma\"ica Clavel}
\affiliation{Univ. Grenoble Alpes, CNRS, IPAG, 38000 Grenoble, France}

\author[0000-0001-6803-2138]{Daryl Haggard}
\affiliation{Department of Physics, McGill University, 3600 University Street, Montreal, QC H3A 2A7, Canada}
\affiliation{Trottier Space Institute at McGill, 3550 University Street, Montreal, QC H3A 2T8, Canada}




\begin{abstract}

Millisecond pulsars (MSPs) are abundant in globular clusters (GCs), which offer favorable environments for their creation. While the advent of recent, powerful facilities led to a rapid increase in MSP discoveries in GCs through pulsation searches, detection biases persist. In this work, we investigate the ability of current and future detections in GCs to constrain the parameters of the MSP population in GCs through a careful study of their luminosity function. Parameters of interest are the number of MSPs hosted by a GC, as well as the mean and the width of their luminosity function, which are typically affected by large uncertainties. While, as we show, likelihood-based studies can lead to ill-behaved posterior on the size of the MSP population, we introduce a novel, likelihood-free analysis, based on Marginal Neural Ratio Estimation, which consistently produces well-behaved posteriors. We focus on the GC Terzan 5, which currently counts 48 detected MSPs. We find that about 158 MSPs should be hosted in this GC, but the uncertainty on this number remains large. We explore the performance of our new method on simulated Terzan 5-like datasets mimicking possible future observational outcomes. We find that significant improvement on the posteriors can be obtained by adding a reliable measurement of the diffuse radio emission of the GC to the analysis or by improving the detection threshold of current radio pulsation surveys by at least a factor two.

\end{abstract}



\section{Introduction}

Globular clusters (GCs) contain about half of the known population of Galactic millisecond pulsars \citep[MSPs,][]{2023ApJ...958..191S}. Recently, the number of MSPs detected in GCs has risen rapidly thanks to new and powerful radio facilities like MeerKAT \citep{2021MNRAS.504.1407R, 2022A&A...664A..27R} and FAST \citep{2021ApJ...915L..28P}. Undoubtedly, MSPs are efficiently produced in GCs, which inform us about their possible formation channels. In addition, GCs have peculiar properties, namely a high stellar density and a profusion of old stars, which could enhance MSP formation \citep{2019ApJ...877..122Y}. Some formation scenarios suggest that MSPs are old pulsars spun up through accretion in binary systems, while others favor former white dwarfs having undergone accretion-induced collapse. These different formation channels for MSPs also impact their abundances. However, to date, the number of MSPs in GCs remains highly uncertain. An alternative way to tackle the problem of their abundance is to study the luminosity function of MSPs in GCs, which is also impacted by their formation history. 

The observation of MSPs in GCs, as well as elsewhere in the Galaxy, suffers from several biases. Brighter MSPs are easier to detect, and the period, dispersion, scattering or scintillation of their pulsed emission also impact their detection \citep{2017MNRAS.472.1458D}. As a result, data only reveal the most luminous sources, likely in an incomplete fashion. Nonetheless, as the performance of radio instruments and their associated data sets improve, it will become easier to accurately extrapolate the luminosity function below the sensitivity of the observations.

Several works have investigated the luminosity function of MSPs in GCs in the past. \cite{2011MNRAS.418..477B} used data from 85 MSPs in 10 GCs and, assuming a common luminosity function for all GCs, found that a log-normal distribution provides statistically better agreement with the data than the other distributions they tested. Their analysis relied on a comparison of Monte-Carlo simulated MSP luminosities with observed ones by performing Kolmogorov-Smirnov and $\chi^2$ tests. The analysis of \cite{2011MNRAS.418..477B} relies on Approximate Bayesian Computation \citep[ABC,][]{2018arXiv180209720S}, although the authors do not refer explicitly to this methodology. ABC tackles inference problems without a likelihood and estimates parameters by measuring a defined distance between simulated and real data. ABC belongs to the larger category of Simulation-Based Inference \cite[SBI,][]{2020PNAS..11730055C}, which has also been dubbed \textit{likelihood-free} inference, as it does not rely on an explicit likelihood but on a data simulator instead. More recently, \cite{2013MNRAS.431..874C} presented a \textit{likelihood-based} Bayesian inference of the parameters of the MSP luminosity function, assuming again a log-normal distribution. Unlike \cite{2011MNRAS.418..477B}, each GC was treated independently from the others, ultimately producing different luminosity functions.

Since the publication of these studies, new MSP detections have been reported, and over the last decade inference techniques have evolved. In particular, ongoing efforts at the intersection of machine learning and SBI have produced new tools \citep{2020PNAS..11730055C} that have not yet been applied to the problem tackled by \cite{2011MNRAS.418..477B} and \cite{2013MNRAS.431..874C}. The goal of this paper is to develop an up-to-date method to robustly determine the number of MSPs in GCs, along with its uncertainty, from individual detections using recently acquired data sets, and a new analysis framework: Marginal Neural Ratio Estimation (MNRE). Traditional inference methods, like the one mentioned above, produce a full, multi-dimensional, joint posterior, which is often not what one actually wants to study. Instead, marginalized and two-dimensional joint posteriors for specific parameters of interest are usually more interesting. Therefore, a large amount of computational time may be spent on solving the full problem while the scientific interest rather lies on a small subset of parameters. Marginal inference estimates the marginal posteriors of interest directly. The goals of the present work is to investigate the performance of SBI in comparison with traditional likelihood-based Bayesian analysis techniques applied to the same physical problem. Our target is to infer from synthetic radio MSP populations what quantities are essential to robustly constrain the population parameters in GCs and what properties help to obtain narrower, and yet well-behaved posteriors, of the inferred parameters. 

In Section  \ref{sec:data_set}, we present the data sets, real and simulated, used in our analysis. Section \ref{sec:bayes_analysis} is dedicated to a discussion of the Bayesian analysis of \cite{2013MNRAS.431..874C}. In Sections \ref{sec:swyft_impl} and \ref{sec:swyft_res}, we present our SBI framework and its results. Discussion and conclusions are presented in Sections \ref{sec:discussion} and \ref{sec:conclusions}, respectively.

\section{Data set}
\label{sec:data_set}

\subsection{Real data set}

The basis for a statistical analysis of the intrinsic MSP population luminosity function 
is the measurement of the flux density, $S$, of a large sample of MSPs in GCs.

\subsubsection{Pulsar detection}
\label{sec:psr_det}

In general, detection of pulsars relies on the measurement of a pulsed emission, not always but usually in the radio domain. A well-identified pulsation period is the key point of a detection, while a flux measurement is not essential and requires calibration beforehand. Indeed, when radio telescopes are used for pulsation searches, they record the \textit{relative} intensity of the sky in the entire field of view of the telescope at a high sampling frequency. As a result, pulsars may lack absolute flux measurements despite their detection, and absolute flux measurements can vary from one instrument to another, see for example, the factor $\sim 2$ difference in the fluxes of Terzan 5 pulsars quoted by \cite{2011MNRAS.418..477B} and \cite{2013MNRAS.431..874C}. Finally, as pulsation searches only record one-pixel images of the sky, the exact position of a pulsar cannot be deduced from a single observation, but requires tracking over several months. Hence, it is not always possible to associate a pulsar with a source seen in a radio image.

To be detected, the flux of a pulsar must lie above a certain flux threshold, which, in general, depends on the conditions along the line of sight as well as the radio telescope used for the observation. This threshold can be quantitatively stated via the radiometer equation \citep{1985ApJ...294L..25D}, which also depends on the pulsation period $P$ of the pulsar:
\begin{equation}
    S_\mathrm{th}(P) \propto \sqrt{\frac{w_\mathrm{obs}}{P - w_\mathrm{obs}}},
\end{equation}
where $w_\mathrm{obs}$ is the observed width of the pulse. A generic feature of the radiometer equation is that it predicts no universal detection threshold but a period-dependence which flattens out towards long periods. Another parameter that strongly impacts the detection threshold is the column density of free electrons, ordinarily referred to as the dispersion measure, which increases the value of $w_\mathrm{obs}$.
For these reasons, an ideal pulsar data set for statistical analyses should ideally come from observations made with a single instrument in a unique configuration, for which both new discoveries and re-detections (of known objects) should be reported. Despite ongoing efforts to collect data towards GCs through uniform, large surveys, see \textit{e.g.}~\cite{2021MNRAS.504.1407R}, a census made with MeerKAT, it does not exist, to the best of our knowledge, a, publicly available, extended list of (re)detected pulsars in GCs quoting fluxes and telescope's observing parameters. In what follows, we will therefore focus on a single GCs, Terzan 5, for which recent, uniform flux measurements, 
have been released.

\subsubsection{Terzan 5}
\label{sec:ter5-real-sample}

Terzan 5, or Ter 5, is the GC with the largest number of identified pulsars, including 49 with detected radio pulsations\footnote{https://www3.mpifr-bonn.mpg.de/staff/pfreire/GCpsr.html}. Forty eight of these have a period P $\leq$ 30 ms and are therefore considered MSPs in the context of our work. The remaining pulsar, Terzan 5 J, has a period P $\simeq$ 80 ms. The decade-old analysis of \cite{2013MNRAS.431..874C} included 25 Terzan 5 pulsars (24 MSPs + Terzan 5 J) all with a flux measurement. As the author noted, 34 pulsars were known at the time, but sources without flux measurement were excluded from the analysis.

\cite{2022ApJ...941...22M} provided the most up-to-date flux and spectral index measurements of pulsars in Terzan 5. They identified 32 sources in archival data of the Green Bank Telescope, at 1.4 and 2 GHz, and computed their spectral indices, assuming power-law spectra. 10 additional fluxes were recorded with MeerKAT at 1284 MHz \citep{2024arXiv240317799P}. Thanks to the spectral index measurements, one can rescale the fluxes of \cite{2022ApJ...941...22M} at the MeerKAT observing frequency. The final data set thus contains 42 pulsars with flux measurements, and 7 pulsars without. This dataset is \textit{incomplete}, as not all of the sources detected in Terzan 5 have a flux measurement. When applying our MNRE framework to Ter 5, we will treat the Ter 5 dataset without Terzan 5 J. Appendix \ref{app:flux_dataset} provides the flux densities used throughout this study. 

\cite{2022ApJ...941...22M} concluded that a dozen to more than a hundred additional pulsars are still to be discovered in Terzan 5. These sub-threshold sources should contribute, at least partially, to the residual diffuse radio emission of Terzan 5, \textit{i.e.}~the total radio emission of the cluster minus the radio fluxes from the resolved point-like sources (and any other source of radio emission in the GC). Except for early observational results \citep{2000ApJ...536..865F} -- most likely limited by the resolution of the instruments at that time -- there are no recent, comparable, assessments of the overall diffuse radio emission from known Galactic GCs. Yet, this quantity, as we will see in what follows, can constrain the cumulative emission from sub-threshold MSPs, and thus may help to narrow down the posterior distribution of the total number of sources in a GC.

\subsection{Mock data set}
\label{sec:mock_data_bayes}

Our ultimate goal is to infer the luminosity function of MSPs in GCs. To this end, we assume that the $\log_{10}$ of the (pseudo-)luminosity $L$ of MSPs in GCs follows a normal distribution, as suggested by previous studies \citep{2011MNRAS.418..477B, 2013MNRAS.431..874C}:
\begin{equation}
    f(\log_{10}(L)) = \frac{1}{\sqrt{2\pi}\sigma} \exp{\left(-\frac{1}{2}\left[\frac{\log_{10}(L)-\mu}{\sigma}\right]^ 2\right)}.
    \label{eq:logl_pdf}
\end{equation}
$L$ is expressed in mJy kpc$^2$ in Equation \ref{eq:logl_pdf} and throughout this study. $\mu$ and $\sigma$ are the mean and the width (\textit{i.e.}~the parameters to infer) of the luminosity function. Given the GC distance, fluxes are shifted according to:
\begin{equation}
    \log_{10}(S_i) = \log_{10}(L_i) - 2 \log_{10}(d).
    \label{eq:logl_to_logs}
\end{equation}
$S_i$ and $L_i$ are, respectively, the flux density and luminosity of pulsar $i$, and $d$ is the distance to the cluster. $d$ is expressed in kpc and $S_i$ in mJy. We can define a probability density function $g(\log_{10}(S))$ that gives the probability of a pulsar in the cluster to have a flux $S$. Unlike $f$, this function is not unique and varies from one GC to another.

\begin{table*}[th]
    \centering
    {
    \begin{tabular}{lccccc}
    \Xhline{5\arrayrulewidth}
     & \multicolumn{2}{c}{Bayesian} & \multicolumn{3}{c}{swyft}\\ 
    Parameter & Sampling & Prior range & Sampling & Prior range & Physical prior range \\ 
    (1) & (2) & (3) & (4) & (5) & (6) \\ 
    \hline
    $N$ & uniform & $\left[N_{\rm detected}, N_{\mathrm{max}}\right]$ & log-uniform & $\left[\log_{10}{N_{\rm detected}}, 2.7\right]$ & $\left[N_{\rm detected}, 500\right]$ \\
    $\mu$ & uniform & $\left[-2.0, 0.5\right]$ & uniform & $\left[-2.0, 0.5\right]$ & $\left[10, 3162\right]$ $\mu$Jy\\
    $\sigma$ & uniform & $\left[0.2, 1.4\right]$ & uniform & $\left[0.2, 1.4\right]$ & -- \\
    $d$ & normal & -- & -- & -- & -- \\
    $S_{\mathrm{th}, \infty}$ & uniform & $\left[0,\min({S_i})\right]$ & log-uniform & $\left[0.5, 1.6\right]$ & $\left[3, 40\right]$ $\mu$Jy\\
    \hline
    \end{tabular}
    \caption{Summary of the applied prior ranges for the Bayesian analysis (Section \ref{sec:bayes_analysis}) and the \sbi-simulator (Section \ref{sec:swyft_impl}) of the GC MSP population. The priors used in the Bayesian analysis are the ones used by \cite{2013MNRAS.431..874C}. The sixth column states the prior ranges of the fifth column in physical units while the fifth column is the numerical input for the respective probability distribution functions.
    \label{tab:swyft_priors}}}
\end{table*}

Considering $N$ the \textit{total} size of the MSP population in the cluster, we simulate the MSP population in the cluster by drawing $N$ mock values of $\log_{10}(L)$ from Equation \ref{eq:logl_pdf}, and then we use Equation \ref{eq:logl_to_logs} to obtain, in turn, $N$ mock values of $\log_{10}(S)$. Next, we identify the mock fluxes larger than a certain detection threshold value $S_{\mathrm{th},i}$ and tag them as \textit{detectable}. $S_{\mathrm{th},i}$ may be different for each MSP (and is generally dependent on the cluster), and varies according to the radiometer equation. In the following, we consider two cases:
\begin{itemize}
    \item[1.] We assume that the detection threshold is cluster- but not MSP-dependent, as assumed by \cite{2011MNRAS.418..477B} and \cite{2013MNRAS.431..874C}.
    \item[2.] We mimic the effect of the radiometer equation by randomly drawing $S_{\mathrm{th},i}$ from a half-normal distribution with mean value $S_{\mathrm{th}, \infty}$ representing the threshold for long-period pulsars. The width of the half-normal is taken to be $\sigma_{\mathrm{th}} = S_{\mathrm{th}, \infty}$ such that larger detection thresholds are possible, accounting for possible systematical errors in the determination of the flux threshold from the radiometer equation, and mimicking an MSP-dependent threshold.
\end{itemize}
For a cluster, the number of \textit{detectable} mock MSPs is $N_\mathrm{detected} \leq N$. MSPs below the (case-by-case) threshold are counted as sub-threshold sources and their cumulative flux $S_{\mathrm{sub}}$ is computed. Together with the flux of detectable MSPs $S_{\mathrm{det}}$, $S_{\mathrm{sub}}$ is responsible for the MSP radio emission of the cluster, $S_{\mathrm{MSP}}$, which is at most equal to the total radio emission of the cluster $S_{\mathrm{tot}}$:
\begin{equation}
\begin{split}
    S_{\mathrm{tot}} \geq S_{\mathrm{MSP}} &= \sum_i S_i = S_{\mathrm{det}} + S_{\mathrm{sub}}\\
    &= \sum_{S_i \geq S_{\mathrm{th}},i} S_i + \sum_{S_i < S_{\mathrm{th}},i} S_i.
\end{split}
\end{equation}
In ideal cases, the value of each $S_i \geq S_{\mathrm{th},i}$ should be known. However, as mentioned in Section \ref{sec:psr_det}, this might not always be the case, and may potentially hamper our ability to infer the posterior distributions of the MSP population. To resemble actual data, we introduce a parameter $p_{\mathrm{fluxless}}$ which represents the proportion of MSPs for which we have a detection but no flux measurement. $p_{\mathrm{fluxless}} = 1 - m / N_{\rm detected} \leq 1$, where $m$ is the number of detected MSPs which have a flux measurement. In our simulation, we guarantee that the fraction of fluxless MSPs matches the one in real data. Finally, we define:
\begin{equation}
    S_{\mathrm{diff}} = S_\mathrm{tot} - \sum_{m} S_i
    \label{eq:Sdiff}
\end{equation}
which corresponds to the diffuse radio flux of the GC. $S_{\mathrm{diff}}$ may originate also from sources other than MSPs so that $S_{\mathrm{diff}} - S_{\mathrm{sub}} \geq 0$. The diffuse flux measurement is nothing but an upper limit on the total flux from the unresolved part of the GC's MSP population.

Our mock datasets realizations are based on the Terzan 5 best-fit results of \cite{2013MNRAS.431..874C}, as we detail below. 

\section{Bayesian analysis}
\label{sec:bayes_analysis}

\subsection{Framework}
\label{sec:bayes_framework}

Our Bayesian analysis relies on the framework described in \cite{2013MNRAS.431..874C} which is briefly summarized here. Let $M$ be our model, $\bm{\theta}$ its parameters and $D$ the data. Then, according to Bayes' theorem:
\begin{equation}
    P(\bm{\theta}|D,M) \propto \mathcal{L}(D|\bm{\theta},M)p(\bm{\theta}|M),
    \label{eq:bayes}
\end{equation}
where $P(\bm{\theta}|D,M)$ is the posterior distribution of the parameters given the data and the model, $\mathcal{L}(D|\bm{\theta},M)$ is the likelihood of the data given the parameters and the model, and $p(\bm{\theta}|M)$ is the prior distribution of the parameters given the model. The proportionality sign accounts for the evidence $p(D)$ that divides the right-hand side of the equation, which, as a constant, can be ignored here. The likelihood $\mathcal{L}(D|\bm{\theta},M)$, associated with a given cluster, takes contributions from three independent likelihoods, namely:
\begin{itemize}
    \item[1.] the likelihood of having a set of pulsars with fluxes $\{S_i\}$, computed as the product of the $g(\log_{10}(S_i))$;
    \item[2.] the likelihood of observing $N_\mathrm{detected}$ (or $m$) pulsars in a cluster with $N$ pulsars, computed as a binomial distribution;
    \item[3.] the likelihood of observing a total flux $S_{\mathrm{tot}}$, which has a Gaussian distribution around $N \langle S \rangle$ where $\langle S \rangle$ is the average MSP flux in the cluster. Here, as done in \cite{2013MNRAS.431..874C}, we assume $S_{\mathrm{tot}} = S_{\mathrm{MSP}}$.
\end{itemize}

Our luminosity model then has 5 parameters: $\bm{\theta} = \{N, \mu, \sigma, S_{\mathrm{th}}, d\}$. We adopt uniform priors for the first 4 parameters. A Gaussian prior is chosen for $d$, reflecting the idea that the distance to the GC is already known with some uncertainty. All priors are independent and similar to the ones used by \cite{2013MNRAS.431..874C}. Table \ref{tab:swyft_priors} summarizes the prior sampling and ranges. Focusing on Terzan 5, the maximal number of MSPs in the cluster is set to 500, and the mean and the width of the distance prior are 5.5 and 0.9 kpc, respectively, according to the result of \cite{2007A&A...470.1043O}.

\subsection{Results}
\label{sec:bayes_res}

\begin{figure}[t]
    \centering
    \includegraphics[width = 0.48\textwidth]{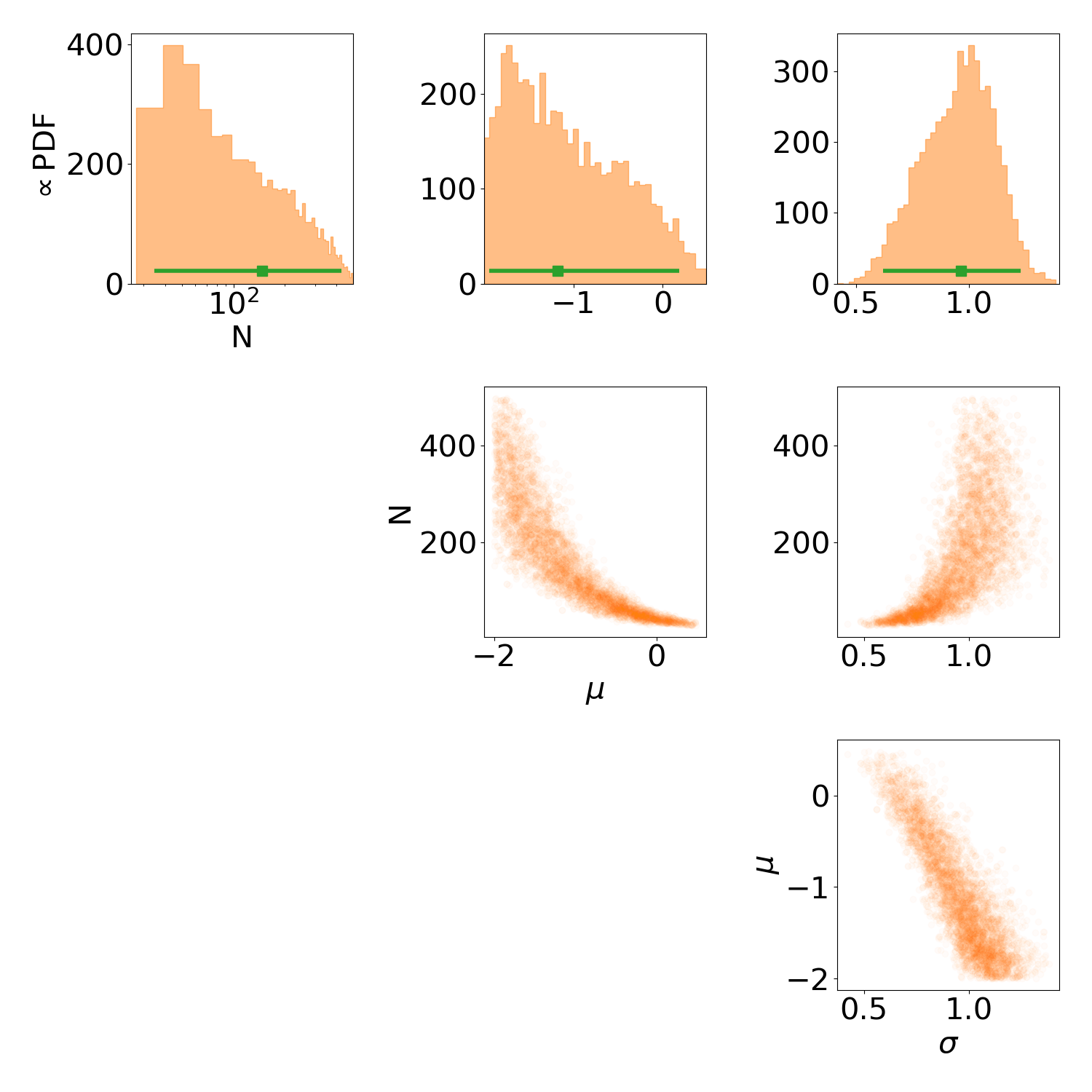}
    \caption{Reproduction of the likelihood-based analysis results of \cite{2013MNRAS.431..874C} for Terzan 5. The top row shows the marginal probability density functions (PDFs) in orange and the median and its 95\% error in green for each parameter. The position of the samples in each 2D parameter space are also shown (second to last rows).}
    \label{fig:ter5_bayes_corner}
\end{figure}

We perform Bayesian analysis using \texttt{PyMultiNest}, a Bayesian inference tool written for \texttt{Python} \citep{2014A&A...564A.125B} which samples the parameter space using a Monte-Carlo algorithm based on nested sampling. While this method samples parameter space more efficiently compared to other methods, the computation time still increases with the number of live points $n_\mathrm{lp}$, \textit{i.e.}, the size of the set of samples drawn from each prior. To identify the number of live points needed for our analysis, we first reproduce the results of \cite{2013MNRAS.431..874C}. We strictly follow their analysis framework, summarized in Section  \ref{sec:bayes_framework}, and apply it to the Terzan 5 pulsar sample, see Section~\ref{sec:ter5-real-sample}. For a meaningful comparison, we used the fluxes of the 25 Terzan 5 MSPs in Table 1 of \cite{2013MNRAS.431..874C}. We conclude that a number of live points $n_\mathrm{lp} \sim 1000$ suits our analysis. Our posteriors, shown in Figure \ref{fig:ter5_bayes_corner}, clearly illustrate that $N$, $\mu$ and $\sigma$ are degenerate to some extent, as already noted by \cite{2013MNRAS.431..874C}.

The 95\% credible interval on $N$ obtained through the likelihood-based analysis is not significantly narrower than the prior chosen for the parameter (\textit{e.g.}~32--452 vs. 25--500). This can indicate that the data are not informative enough to constrain the luminosity function and the size of the MSP population in Terzan 5. 

This hypothesis can be tested by assessing the coverage of the credible intervals: In case of correct coverage, the $x$\% credible interval of a parameter should contain the true parameter value in $x$\% percent of the cases. If the $x$\% credible region contains the true parameter value in $y$\% percent of the cases, where $y>x$, the credible interval is too wide and it is said to be \textit{conservative}. On the other hand, if $y<x$, the interval is too narrow and it is said to be \textit{over-confident}. Our goal in this section  is to assess the quality of the coverage obtained with the Bayesian analysis, which ultimately inform us about the correctness of the statistical inference. If the data are genuinely not informative enough to constrain the number of MSPs in Terzan 5, we should find $x\simeq y$ for all values of $x$.
We simulate 500 mock data sets according to Section  \ref{sec:mock_data_bayes} using as true parameters of the model some of the results obtained by \cite{2013MNRAS.431..874C} for Terzan 5. Namely, we assume $N = 142$, $\mu = -1.2$ and $\sigma = 1.0$. The sensitivity threshold is set to $S_\mathrm{\mathrm{th}} = 0.02$ mJy and the distance to $d = 5.5$ kpc. $N_\mathrm{detected}$ varies from one simulation to another. Finally, as in \cite{2013MNRAS.431..874C}, we have $p_\mathrm{fluxless} = 0$, \textit{i.e}, $m = N_\mathrm{detected}$. We apply the Bayesian analysis streamlined in Section  \ref{sec:bayes_analysis} to the mock data sets and use the posteriors of $N$ to assess the coverage of this parameter. We build a symmetric interval around the median $N_\mathrm{med}$ of the posterior distribution $P(N)$ such that the integral of the posterior from the lower bound of the interval $N_\mathrm{low}$ to the median equals the integral of the posterior from the median to the upper bound of the interval $N_\mathrm{high}$. We start from $N_\mathrm{low} = N_\mathrm{high} = N_\mathrm{med}$ and increase the size of the interval until $N_\mathrm{low} = 142$ or $N_\mathrm{high} = 142$. We compute:
\begin{equation}
\begin{split}
    x &= 100 \times \int_{N_\mathrm{low}}^{N_\mathrm{high}} P(N) \mathrm{d}N\\
    &= 200 \times |C(142)-C(N_\mathrm{med})|
\end{split}
\end{equation}
where $C$ is the cumulative distribution function associated with $P(N)$. When the true value is close to (far from) the median value, $x$ takes a low (high) value. By definition, $x$ cannot be larger than 100. If $x_k$ is the value that $x$ takes in simulation $k$, $y$ takes the values  $y_k$ computed as the percentage of values of $y$ among all simulations that verify $y\leq y_k$. Hence, the $y_k$ reflects the cumulative distribution of the $x_k$. The values taken by $x$ and $y$ are converted from percentage to significance level, which is the inverse of the error function of the percentage divided by square root of two. This conversion is convenient for plotting and interpreting the coverage. A perfectly calibrated posterior follows a diagonal in the space of \textsc{nominal} versus \textsc{empirically} derived coverage. While this coverage test is indicative of ill-calibrated posteriors that necessitate a re-design of the inference pipeline, a positive outcome for this test does not imply that the obtained posterior distributions are optimal given the available statistical power of the dataset. 

\begin{figure}[t]
    \centering
    \includegraphics[width = 0.4\textwidth]{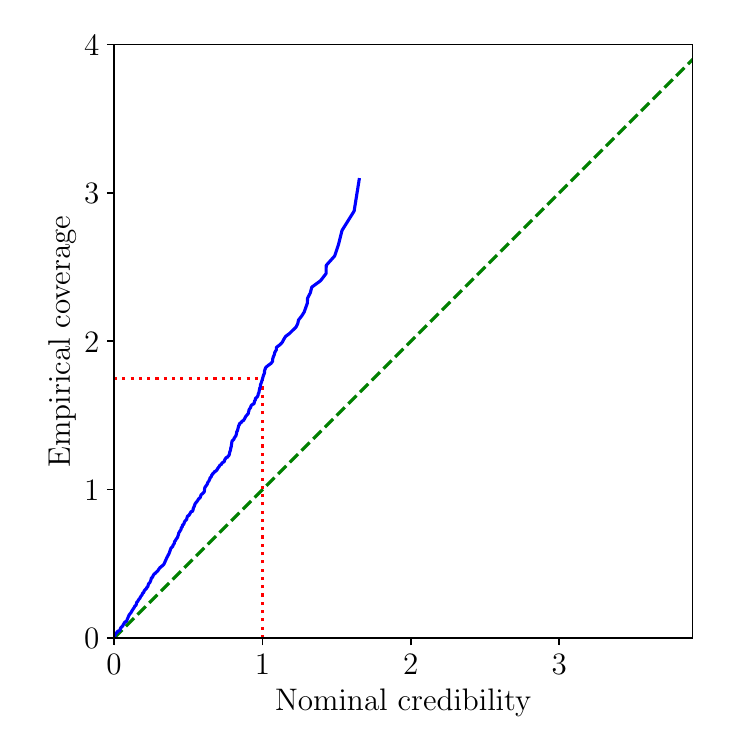}
    \caption{Coverage (blue line) of $N$ obtained from the application of the Bayesian analysis to 500 mock data sets ($N=142$, $\mu=-1.2$ and $\sigma=1.0$). Ideally, the coverage line should follow the green dashed diagonal line. Here, the plot tells us that the method produces conservative results, \textit{i.e.} that the posteriors are too wide. The red lines illustrate that the 68\% (1 $\sigma$) confidence interval contains the true value in more than 68\% of the cases.}
    \label{fig:ter5_bayes_coverage}
\end{figure}

Our results, shown in Figure \ref{fig:ter5_bayes_coverage}, indicate that the posteriors of $N$ tend to be too conservative. For all of them, the true value of $N =142$ falls within the 92.7\% credible interval (1.79 confidence level), while this should only happen for 92.7\%, \textit{i.e.}, 463, of them. The coverage at higher confidence level cannot be estimated because the true value never falls within the 99\% credible interval but outside of the 92.7\% confidence interval. Therefore, a different analysis can obtain a narrower posteriors with the actual data, while still being statistically correct. We note that the general aspect of the coverage may depend on the values chosen for the true parameters. What our results tell us is, should the actual parameters of Terzan 5 be the one we chose (see also Section \ref{sec:swyft_res}), there is a high chance that the Bayesian analysis would yield posteriors that are too wide for the number of sources in the cluster.

\section{Swyft implementation on MNRE}
\label{sec:swyft_impl}

Marginal ratio estimation focuses on the likelihood-to-evidence ratio, which, according to Bayes' theorem, equals the posterior-to-prior ratio:
\begin{equation}
\begin{split}
    r(\bm{\theta},D,M) &= \frac{\mathcal{L}(D|\bm{\theta},M)}{p(D)} = \frac{p(\bm{\theta}|D,M)}{p(\bm{\theta}|M)}\\
    &= \frac{p(D,\bm{\theta})}{p(D)p(\bm{\theta})}.
\end{split}
\end{equation}
Let us now define a binary variable $Y$, such that $Y=1$ when pairs $(D,\bm{\theta})$ are jointly drawn from $p(D,\bm{\theta})$, and $Y=0$ when pairs are marginally drawn from $p(D)p(\bm{\theta})$. One can show that training a binary classifier $f_\phi$ for $Y$ is equivalent to learning the ratio $r(\bm{\theta},D,M)$.

We rely on a specific implementation of MNRE: \sbi. 
This is a simulation-base inference tool that uses marginal neural ratio estimation methods, \textit{i.e.}, its classifier $f_\phi$ uses a neural network and learns from mock data produced by a simulator \citep{Miller:2021hys, Miller:2022shs}. \sbi~has been proven to be efficient at inferring cosmological parameters from cosmic microwave background measurements with posterior convergence achieved using orders of magnitude fewer calls
than Markov Chain Monte Carlo methods \citep{2022JCAP...09..004C}. More recently, \cite{2023MNRAS.518.2746A} showed that the estimation of the warm dark matter mass from strong lensing images could also benefit from MNRE.

\subsection{Simulator}
\label{sec:swyft_simulator}

\begin{figure*}[p]
    \centering
    \includegraphics[width=0.6\columnwidth]{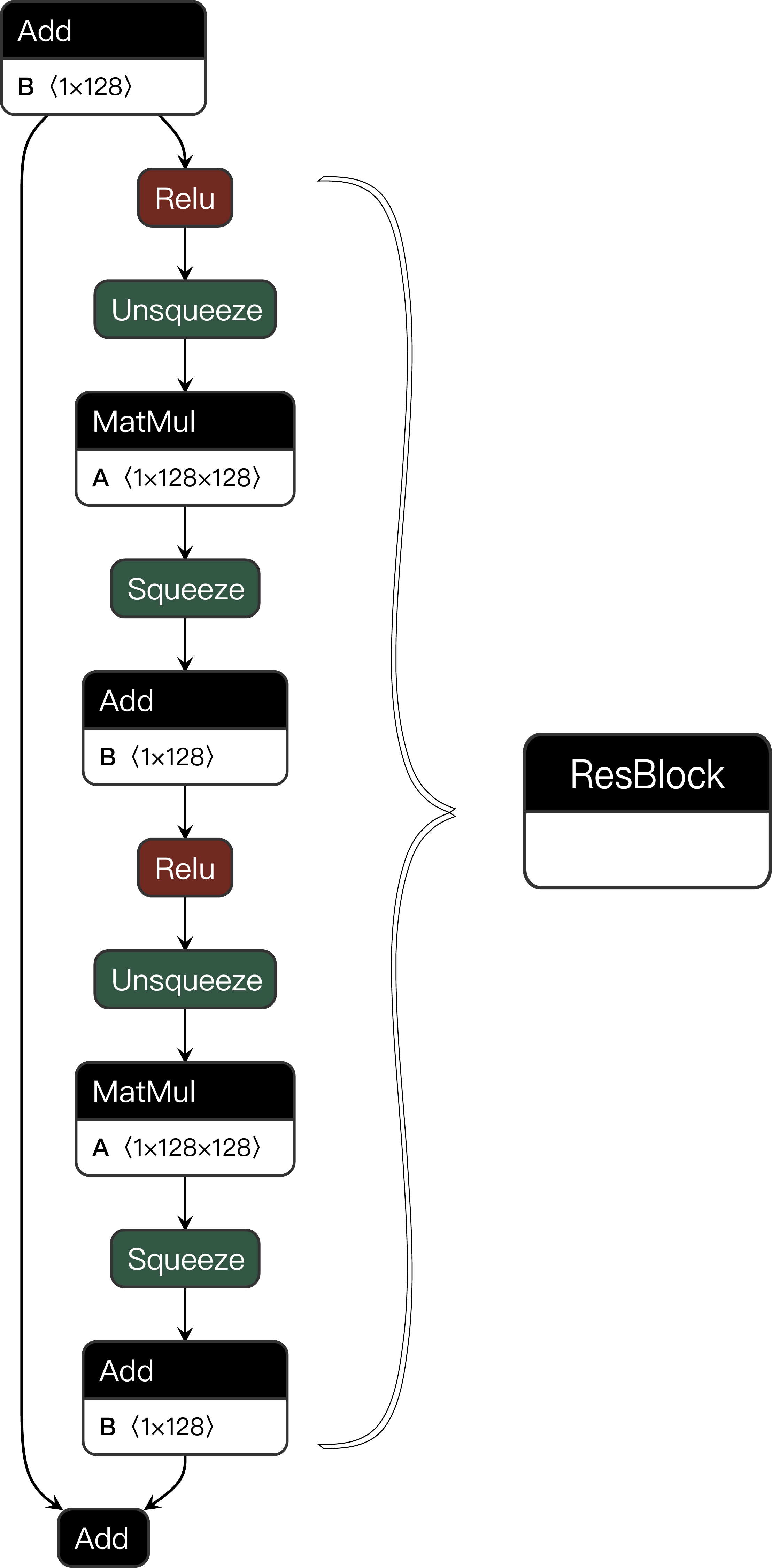}\hspace{3cm}\includegraphics[width=0.29\columnwidth]{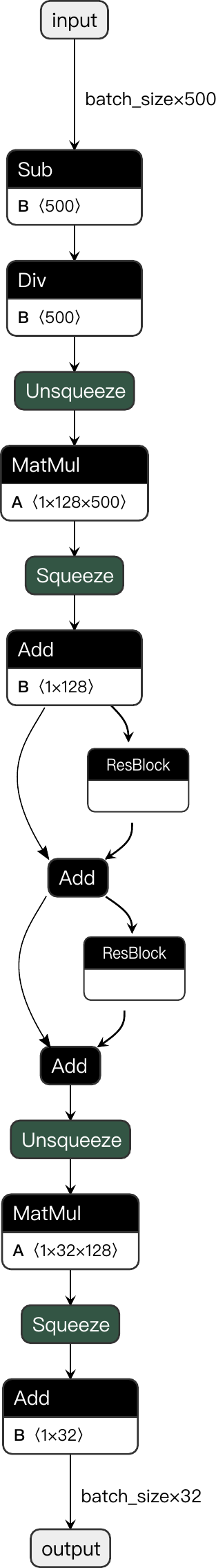}
    \caption{Flow chart of the \texttt{ResNet} implementation used in this work to generate summary statistics for the \sbi~log-ratio estimator. (\emph{Left}:) The general structure of a block ``ResBlock'' of the \texttt{ResNet} used as the core ingredient to the overall \texttt{ResNet} layout. The operation \texttt{Add} performs an element-wise binary addition on a vector of a given size. \texttt{MatMul} refers to a matrix multiplication based on the dimensions specified in the box. \texttt{Unsqueeze} and \texttt{squeeze} augment or reduce the dimension of the input vector. \texttt{Relu} is the standard activation function known as rectified linear function, $y = \max{(0, x)}$, applied element-wise to an input vector. (\emph{Right}:) Full structure of a \texttt{ResNet} defined with two ``ResBlocks'' following the hyper-parameters listed in Table \ref{tab:swyft_resnet} for the case without $S_{\mathrm{diff}}$ summarizing the information in the list of detected MSP fluxes. The operations \texttt{Sub} and \texttt{Div} perform the normalization of the input flux vector via subtraction of the mean and division by the total flux. The other \texttt{ResNet} structures are similar in shape with altered numbers of ``ResBlocks'' and input and output features.}
    \label{fig:flow_chart}
\end{figure*}

In the framework of SBI, we are not bound to limit the number of model parameters and construct a likelihood function in the first place. In particular, the inclusion of nuisance parameters is facilitated. Hence, we extend the mock data generation outlined in Section \ref{sec:mock_data_bayes} to profit from the capabilities of SBI. Our \sbi~radio MSP simulator generates mock data for a single GC. The baseline simulator depends on the parameters described in Section~\ref{sec:mock_data_bayes}, whose priors are varied as reported in Table~\ref{tab:swyft_priors}. We adopt a varying flux threshold, implemented as nuisance parameter, which mimicks an MSP-dependent threshold, Section \ref{sec:mock_data_bayes}.

\textit{Simulator.} This simulator yields a catalog of detected MSPs as well as the additional cumulative flux of the population too dim to be resolved individually. The information about the flux of each detected source is stored in an array, sorted in descending order. The length of the array is set to a fixed number (the upper bound of the prior on $N$). Array entries without a detected source are assigned the fill value ``-1''. As we can have $p_{\mathrm{fluxless}} \neq 0$, the flux value is replaced by the fill value ``-1'' for above-threshold, fluxless, sources. The simulator generates a second array that states $N_\mathrm{detected}$ subtracted by the cumulative sum of detected sources. 

\textit{Prior choices.} We aim to infer the posterior distributions for the model parameters based on the set of priors summarized in Table \ref{tab:swyft_priors}. These priors are slightly wider than what was used in \cite{2013MNRAS.431..874C} and in Section  \ref{sec:bayes_analysis} except for the prior on $\sigma$, which we adopt from this earlier work. Previous scans of the luminosity functions' parameter ranges seem to support this prior range \citep{2011MNRAS.418..477B}. We note that our objective is to infer information from simulated data so that we are rather agnostic about the frequency at which we sample the luminosity function and at which we define the detection threshold. Note that in contrast to Section \ref{sec:bayes_analysis}, we use a log-uniform prior on the number of MSPs $N$. We made this change to more evenly sample the larger range of up to 3000 sources instead of 500. When we later analyze the real sample of Terzan 5 (see Section \ref{sec:results-swyft-Ter5}), we will adopt the same prior definition. 

\subsection{Architecture}
\label{sec:swyft-architecture}

As described in Section \ref{sec:swyft_impl}, the parameter inference in neural ratio estimation relies on training a classifier $f_{\phi}$ that allows for discriminating samples drawn from a joint or a marginal probability distribution. This classifier is a neural network in the case of \sbi.
Hence, design decisions for the network are heavily influencing the robustness, performance and quality of the inference. We have to choose a network architecture that is suitable for the problem at hand and adequate for the data format generated by the simulator which itself is motivated by the structure of real observations of radio MSPs. In our case, we obtain a catalog of detected sources represented by a two-dimensional array as detailed in Section \ref{sec:swyft_simulator}; in abstract terms, a matrix with a certain number of rows and columns. 

\textit{Inference network architecture}. This data structure allows us to resort to network architectures typically employed in image processing. To increase the flexibility of the network, we choose the deep neural network architecture of the \texttt{ResNet} \citep{2015arXiv151203385H}, which has been used in the somewhat similar context of source detection in gamma-ray data of the \textit{Fermi} Large Area Telescope \citep{AnauMontel:2022ppb, Horangic:2023eju}. In particular, we utilize the \texttt{ResNet} implementation natively provided in \sbi.  Each column of the data array is first passed through a normalizing layer and afterwards processed by a dedicated \texttt{ResNet} (with different structural properties) that takes one-dimensional vectors as input. The chosen \texttt{ResNet} hyper-parameters appear in Table \ref{tab:swyft_resnet}. In Figure \ref{fig:flow_chart}, we provide a visualization of the \texttt{ResNet}'s core structure. The output vectors of each \texttt{ResNet} are concatenated to a single vector, which is subsequently passed to the default ratio estimator implementation of \sbi. Thus, the initial data processing with \texttt{ResNet}s compresses the data and generates a summary statistic learned and optimized during the network's training.

\textit{\sbi~hyper-parameter settings.} Besides the structural hyper-parameters of the \texttt{ResNet}, training and inference with the MNRE algorithm of \sbi~requires the specification of further parameters. A summary of the parameter choices is provided in Table \ref{tab:swyft_resnet}. A subset of the stated parameters controls the training procedure: The sample size refers to the total number of simulated radio MSP populations for the training. This sample is split into a subset used for training (70\%) and a validation dataset (30\%) used to evaluate the performance of the trained network on data it has not seen during the training iterations. Both datasets are fed into the network in batches of 64 samples. The network training is performed with the Adam optimizer in, at most, 100 epochs starting with an initial learning rate of $8.5\times10^{-4}$. This learning rate may decrease during the training if, after five consecutive epochs (learning ratio schedule patience), no improvement in the training loss was achieved. In this case, the current learning rate is reduced by 30\%. If 20 consecutive epochs did not result in an improvement in the training loss (early stopping patience), the training terminates immediately. Lastly, we allow for noise resampling during the training process. The idea is to re-generate the noise in the data while keeping the training samples the same. In practice, the noise is represented by the set of detected sources without flux measurements. For each sample, we repeat per epoch the selection of MSPs that are listed with a flux value as the simulator has stored the full catalog of detected sources with their original randomly drawn fluxes. This procedure effectively enhances the number of training samples since the dataset never looks like the one used in the previous epoch. It has been shown that noise resampling is a crucial way to prevent overfitting the training data and it stabilizes the shape of the posterior distributions \citep{Alvey:2023pkx, Alvey:2023naa}. 

\begin{table*}
    \centering
    {\renewcommand{\arraystretch}{1.5}
    \begin{tabular}{lcc}
    \Xhline{5\arrayrulewidth}
    Hyper-parameter & Value (w/o diffuse flux) & Value (w diffuse flux) \\ 
    \hline
    \texttt{ResNet}: \# input features & (500 / 500) & (500 / 500 / 500) \\
    \texttt{ResNet}: \# output features & (32 / 16) & (32 / 16 / 32) \\
    \texttt{ResNet}: \#  hidden features & (128 / 128) & (128 / 128 / 128) \\
    \texttt{ResNet}: \# blocks & (2 / 4) & (2 / 4 / 5) \\ 
    \hline
    Sample size & \multicolumn{2}{c}{$10^5$} \\
    Training-to-validation ratio & \multicolumn{2}{c}{$70:30$}\\
    Training/validation batch size & \multicolumn{2}{c}{$64/64$} \\
    Optimizer & \multicolumn{2}{c}{Adam}\\
    Initial learning rate & \multicolumn{2}{c}{$8.5\times10^{-4}$}\\
    Learning rate scheduler  & \multicolumn{2}{c}{\texttt{ReduceLROnPlateau}}\\
    Learning ratio schedule decay factor & \multicolumn{2}{c}{0.3}\\
    Learning ratio schedule patience & \multicolumn{2}{c}{5}\\
    Maximal number of training epochs & \multicolumn{2}{c}{$100$}\\
    Early stopping patience & \multicolumn{2}{c}{$20$} \\
    Noise resampling & \multicolumn{2}{c}{\texttt{true}}\\
    \hline
    \end{tabular}
    \caption{Summary of network architecture and \sbi~hyper-parameters used in this work; as outlined in Section \ref{sec:swyft-architecture}. For the \texttt{ResNet} hyper-parameters, the numbers in parentheses denote the values selected for each data column. 
    \label{tab:swyft_resnet}}}
\end{table*}

\textit{Including diffuse flux measurements.} Motivated by the approach in \cite{2013MNRAS.431..874C}, we prepare an extension of the basic formalism outlined above. We aim to explore the impact of an additional diffuse radio flux measurement $S_{\mathrm{diff}}$ by extending the simulator: We create a third column of the generated catalog that utilizes the already computed cumulative flux of sub-threshold sources $S_{\mathrm{sub}}$. All rows with no entry or a sub-threshold source are assigned the value $S_{\mathrm{tot}} - S_{\mathrm{sub}}$. Subsequently following the ordering of flux in the first column, this value is decremented by the flux of the respective detected MSP. Sources without a flux measurement are included with zero flux.

\textit{\sbi~hyper-parameters with diffuse flux.} This scenario does not require much tuning of the hyper-parameters chosen in the baseline setup, but we must add a third \texttt{ResNet} that accepts the list of fluxes describing diffuse and resolved contributions. The selected parameters are stated in the third column of Table \ref{tab:swyft_resnet}.

\section{Swyft results}
\label{sec:swyft_res}

\begin{table*}[t]
    \centering
    \begin{tabular}{lcc}
        \Xhline{5\arrayrulewidth}
        Case & Injected $N$ & Inferred $N$\\
        \hline
        (1) $N_\mathrm{detected}$ = 17 & 142 & 62$^{+298}_{-44}$ \\
        (2) $N_\mathrm{detected}$ = 40 & 200 & 135$^{+279}_{-86}$ \\
        (3) $N_\mathrm{detected}$ = 40, extended prior & 200 & 123$^{+412}_{-79}$ \\
        (4) $N_\mathrm{detected}$ = 40, $S_\mathrm{diff} = 2$ mJy & 200 & 168$^{+200}_{-78}$ \\
        (5) $N_\mathrm{detected}$ = 36, $p_\mathrm{fluxless}= 0\%$ & 200 & 124$^{+313}_{-86}$ \\
        (6) $N_\mathrm{detected}$ = 40, $p_\mathrm{fluxless}= 10\%$ & 200 & 133$^{+277}_{-86}$ \\
        (7) $N_\mathrm{detected}$ = 34, $S_\mathrm{diff} = 2$ mJy, (8) $p_\mathrm{fluxless}= 10\%$ & 200 & 222$^{+229}_{-128}$ \\
        (8) $N_\mathrm{detected}$ = 40, $S_\mathrm{diff} =2$ mJy, $p_\mathrm{fluxless}= 10\%$ & 200 & 185$^{+240}_{-94}$ \\
        (9) $N_\mathrm{detected}$ = 51, $S_\mathrm{diff} =$ 2 mJy, $p_\mathrm{fluxless}= 10\%$ & 200 & 188$^{+201}_{-84}$ \\
        \hline

    \end{tabular}
    \caption{Median and 95\% credible interval of the number of MSPs $N$ obtained through the \sbi~analysis applied to mock data for various cases.}
    \label{tab:nmed_swyft}
\end{table*}

Now that we have established the simulator of radio MSPs as well as the inference approach, we aim to study the performances of the method and the properties we are able to extract given a realization of synthetic target data. The idea is to choose the underlying model parameters to well represent what an observationally obtained radio MSP catalog looks like. 

\subsection{Mock data}
\label{sec:results-singleGC}

We examine two distinct mock MSP populations: First, we make the direct connection to the Bayesian analysis of Section~\ref{sec:bayes_res} and adopt the same defining population parameters. Then, we turn towards a mock data definition that is oriented closer to the current observational data for Ter 5. The latter case serves as a testing ground to explore the capabilities of our simulator and SBI. We investigate different scenarios for the parameters impacting the quality of the inference, {\it i.e.}, the percentage of detected sources without flux measurements, $p_{\mathrm{fluxless}}$, the availability of diffuse measurements, and the detection threshold at large periods $S_{\mathrm{th},\infty}$. For what follows, the \texttt{swyft} analysis is entirely based on the simulator described in Section \ref{sec:swyft_simulator} with priors stated in Table \ref{tab:swyft_priors} and the inference architecture of Section \ref{sec:swyft-architecture} adhering to the specifications in Table \ref{tab:swyft_resnet}.

\subsubsection{Mock population following \cite{2013MNRAS.431..874C}}
\label{sec:swyft-initial-mock-data}

As a reminder, the MSP population in this case is characterized by $N = 142, \mu = -1.2, \sigma = 1.0$, $S_{\mathrm{th}, \infty} = 0.02$ mJy, $p_{\mathrm{fluxless}} = 0\%$. We generate 1000 realizations of this MSP population and determine the mean expected number of detected MSPs (while keeping the size of the population the same). We obtain  $N_\mathrm{detected} = 17$, which we fix as a hyper-parameter of our simulator.

\begin{figure*}[t]
    \begin{centering}
    \includegraphics[width=\columnwidth]{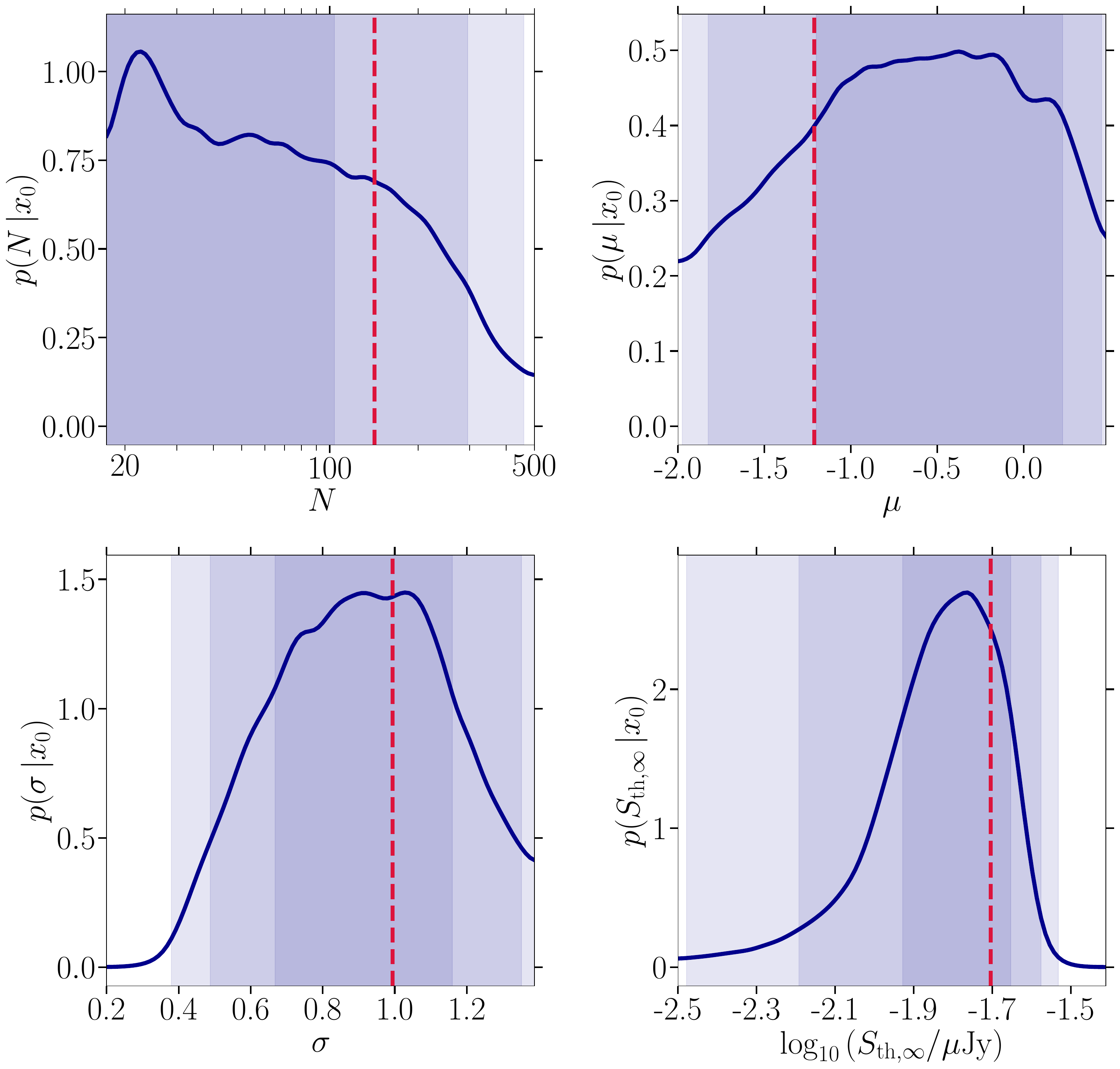}\hfill
     \includegraphics[width=\columnwidth]{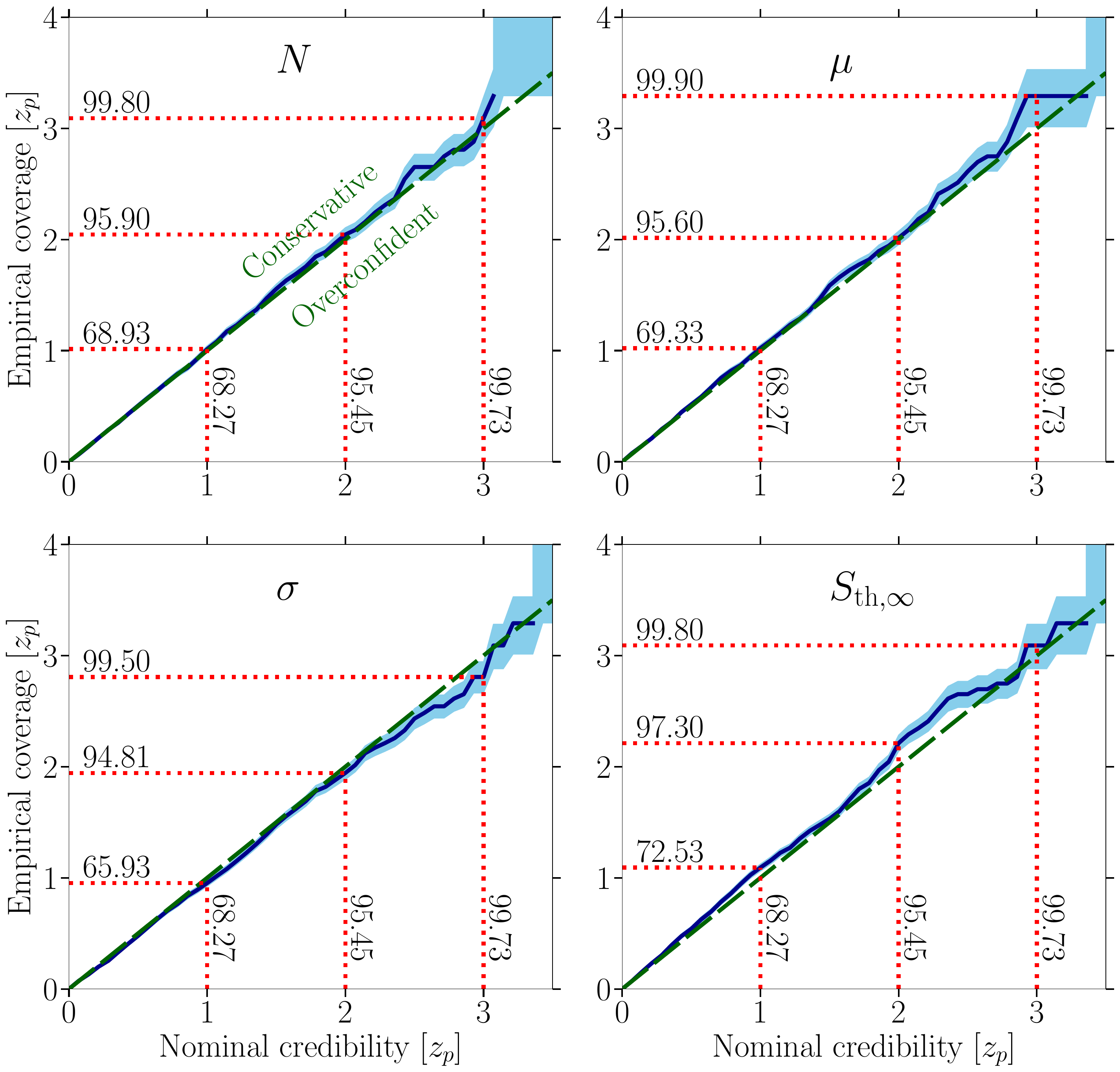}   
    \par\end{centering}
    \caption{(\emph{Left}:) Complete set of one-dimensional marginal posterior distributions (blue curves) for the four parameters characterizing the radio MSP population model similar to the one found in \cite{2013MNRAS.431..874C} with parameters: $N_\mathrm{detected} = 17, N = 142, \mu = -1.2, \sigma = 1.0$, $S_{\mathrm{th}, \infty} = 0.02$ mJy, $p_{\mathrm{fluxless}} = 0\%$. The corresponding $1\sigma$, $2\sigma$, and $3\sigma$ contours are displayed as shaded bands in decreasing opacity. The vertical dashed red lines denote the true parameters of the target observation. (\emph{Right}:) \sbi~coverage results for the mock dataset of the left panel. The horizontal axis states the nominal credibility interval (in significance) while the vertical axis shows the empirically determined coverage. The dashed green line indicates perfect coverage while the solid blue line is the obtained average coverage; the light-blue band denotes the 68\% containment band derived from 1000 simulations.
    \label{fig:mock-swyft-bayesian-comparison}}
\end{figure*}

We visualize the results of the parameter inference with \sbi~in the left panel of Figure~\ref{fig:mock-swyft-bayesian-comparison}. The plot depicts the one-dimensional marginal posterior distributions (blue line) for all four model parameters with corresponding $1\sigma$, $2\sigma$, and $3\sigma$ contours highlighted as shaded bands with decreasing opacity. The injected parameter values are marked with red, dashed lines. The training and subsequent parameter estimation lasted around one and a half hours on a single \texttt{NVIDIA} A100 GPU. A direct comparison with Figure~\ref{fig:ter5_bayes_corner} reveals a striking similarity of the shape of the posterior distributions for $N$ and $\mu$ and $\sigma$. Ultimately, both methods produce almost the same qualitative results, which is reassuring in the sense that the available information is equally well translated into posterior distributions. The median and the 95\% credible of $N$ are reported in the first line of Table \ref{tab:nmed_swyft}.

One may wonder how well the \sbi~posteriors express constraints on the inferred parameters given the information in the datasets. In particular, we are interested in the coverage of the posterior distributions.
Using our MSP population simulator, we generate 1000 mock observations of our selected parameter tuple. We infer with our trained network the posterior distributions for each mock observation and count for how many observations the credible intervals from $x\%\in\left[0, 1\right]$ encompass the true value. We provide the results of this coverage test in the right panel of Figure~\ref{fig:mock-swyft-bayesian-comparison}. 
This derived coverage demonstrates that our \sbi~setup produces reasonably calibrated posterior distributions for all parameters. Some of the coverage profiles are slightly conservative, especially the inference on $S_{\mathrm{th},\infty}$. However, this parameter does not bear critical information about the MSP population since it is technically known from the sample we use. We emphasize that these coverage tests do not make any statement about how well we have made use of the available information in the dataset but rather assess the validity of the posteriors with respect to the inference problem at hand. 
The \sbi~coverage results are in stark contrast to those of the Bayesian analysis in Figure~\ref{fig:ter5_bayes_coverage}. While we demonstrated that both methods produce comparable posterior distributions, their statistical calibration differs substantially. We find that the \sbi~pipeline is a statistically more robust approach to infer the properties of MSP populations in GCs.

Yet, the MSP population parameters chosen here do not align well with our current knowledge of the MSP content of Ter~5 (see Section~\ref{sec:ter5-real-sample}). To better assess the capabilities of the \sbi~approach we switch to a mock data definition that more closely resembles the real Ter~5 dataset. In the following section, we will describe and examine this baseline setup in greater detail.

\subsubsection{Baseline case}
\label{sec:swyft_mock_baseline}

\begin{figure}[t]
    \begin{centering}
    \includegraphics[width=\linewidth]{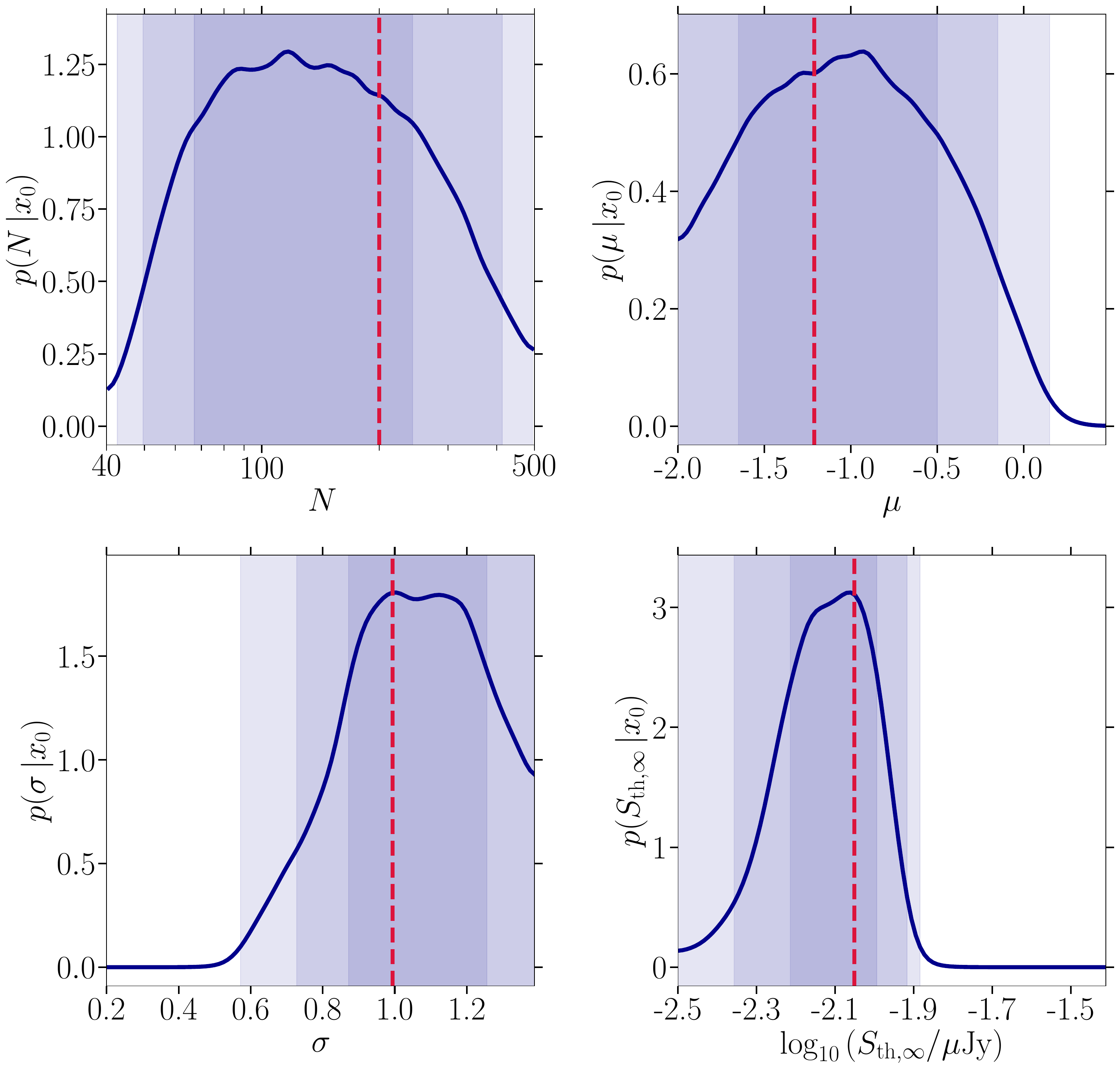}
    \par\end{centering}
    \caption{Same as Figure~\ref{fig:mock-swyft-bayesian-comparison} for our baseline target mock observation with parameters: $N_\mathrm{detected} = 40$, $N = 200$, $\mu = -1.2$, $\sigma = 1.0$ and $S_{\mathrm{th},\infty} = 9$ $\mu$Jy and $p_{\mathrm{fluxless}} = 0\%$.
    \label{fig:singleGC-pflux0p0_wodiff}}
\end{figure}

Ter 5 contains around 40 MSPs with measured fluxes. We adopt this number as $N_\mathrm{detected}$ in what follows. Because of this large number of known MSPs, we increase the total size of our mock MSP population to $N = 200$ while we keep the luminosity parameters as in the previous section, that is, $\mu = -1.2$ and $\sigma = 1.0$. To achieve the targeted number of detected sources with these parameters, we have to set the detection threshold to $S_{\mathrm{th},\infty} = 9$~$\mu$Jy. This value is close to the lowest detected flux of 8 $\mu$Jy of an MSP in Ter 5 \citep{2024arXiv240317799P} at 1284 MHz. Our reported inference results refer to this frequency band, and we notice that our choices for this baseline mock population are reasonable and informed by observations. We start in an ideal setting where all detected MSPs come with a secured flux measurement ($p_{\mathrm{fluxless}} = 0\%$). Finally, we do not include information from the diffuse flux measurement. Our results for this baseline case are illustrated in Figure~\ref{fig:singleGC-pflux0p0_wodiff} following the same style as the left panel of Figure~\ref{fig:mock-swyft-bayesian-comparison}. The median and the 95\% credible intervals of $N$ are reported in the second line of Table \ref{tab:nmed_swyft}. We show the coverage results for this case in Appendix \ref{app:coverage-plots} Figure \ref{fig:coverage-baseline_woDiff}. All the following findings are accompanied by their coverage reported in Appendix \ref{app:coverage-plots}.

We find that the injected values can be recovered with much better precision than in the \cite{2013MNRAS.431..874C} setup; for all four parameters, we are able to recover well-defined credible intervals and not mere upper limits as for the number of sources $N$ in Figure~\ref{fig:mock-swyft-bayesian-comparison}. The detection threshold is reconstructed with the highest precision (and accuracy). This result is very intuitive since the detection threshold is given by the smallest number in the flux column of our synthetic catalogs. We notice, however, that this parameter can already be quite constrained by observations, especially in the case of a uniform survey. More relevant are the MSP population parameters. Here, we find that all injected parameter values are recovered within the $1\sigma$ credible interval of the inferred posteriors. Yet, the posteriors still allow for a wide range of MSP population scenarios, in particular, there might be 50 to 400 MSPs in this mock population at the $2\sigma$ level. A similar statement holds for the viable range of $\mu$.

While our posterior distributions do not run against the prior boundaries, it is interesting to check how robust these inference results are against widening the original priors. As $N$ is not strictly constrained in this baseline setup, we launch an alternative \sbi~run with a larger prior from 40 to 3000 sources while keeping the other priors the same. The median and the 95\% credible regions of $N$ are reported in the third line of Table \ref{tab:nmed_swyft}. The results are shown in the left panel of Figure~\ref{fig:wide-priors-baseline} of Appendix \ref{app:wide-priors}. Inspection of obtained posterior distributions does not reveal major changes to the results shown here in the main text. They are, in fact, almost identical so that we proceed by only considering the prior definitions detailed in Table~\ref{tab:swyft_priors}.

\subsubsection{Impact of diffuse radio emission}
\label{sec:swyft_diffuse}

\begin{figure}[t]
\begin{centering}
\includegraphics[width=\linewidth]{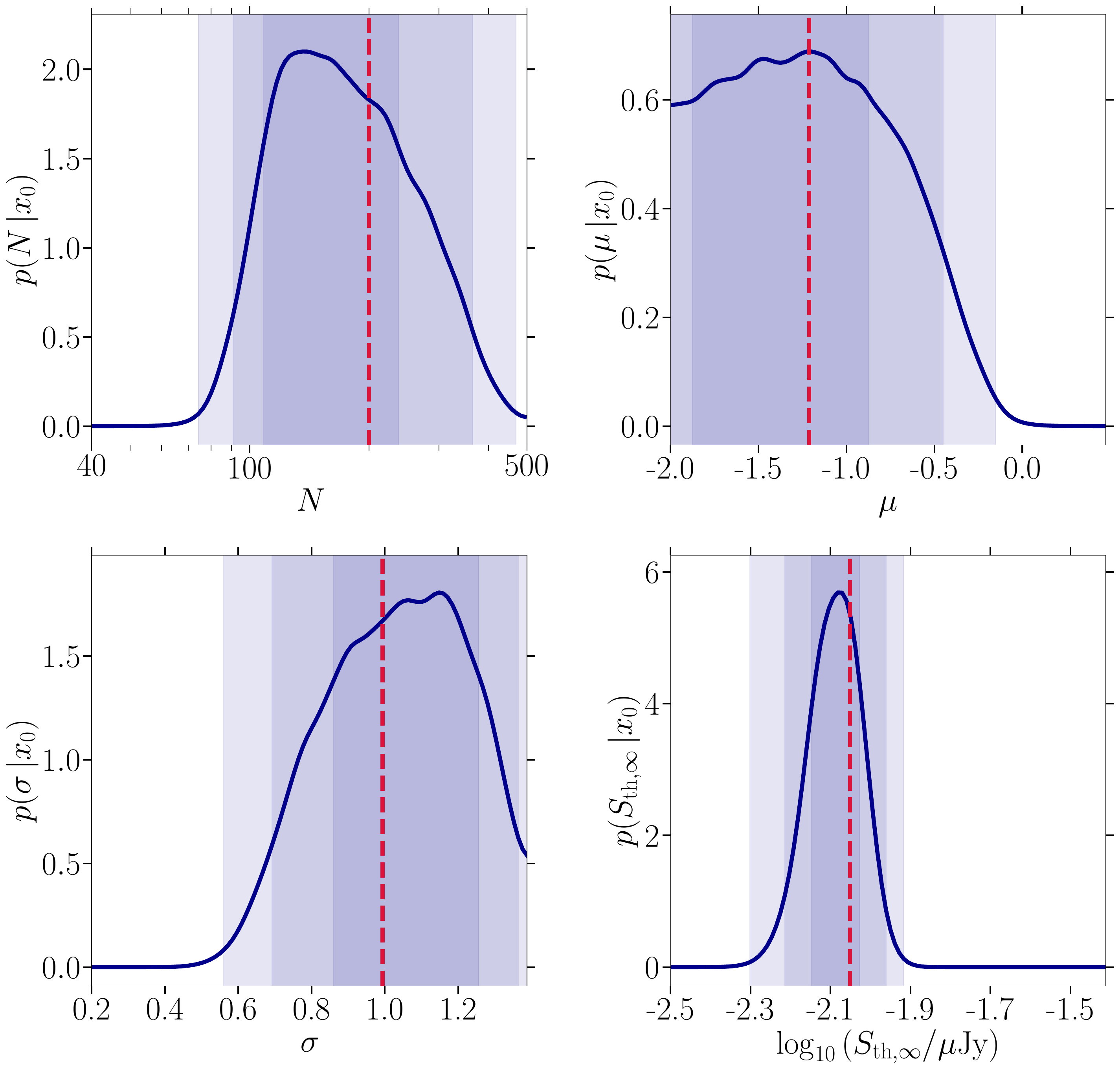}
\par\end{centering}
\caption{Same as Figure \ref{fig:singleGC-pflux0p0_wodiff} including information about the diffuse radio emission in the simulator for which we assume $S_{\mathrm{diff}} = 2$ mJy and $p_{\mathrm{fluxless}} = 0\%$.
\label{fig:singleGC-pflux0p0_wdiff}}
\end{figure}

To include information about the diffuse radio flux, we assume a value of $S_{\mathrm{diff}} = 2$ mJy that adds on top of the cumulative flux of all detected MSPs, as per Equation~\ref{eq:Sdiff}. The results of the parameter estimation are shown in Figure \ref{fig:singleGC-pflux0p0_wdiff} and the median and the 95\% credible intervals of $N$ are reported in the fourth line of Table \ref{tab:nmed_swyft}. The corresponding coverage results are shown in Figure \ref{fig:coverage-baseline-wDiff} of Appendix \ref{app:coverage-plots}.

Compared to our baseline case (Section \ref{sec:swyft_mock_baseline}), adding information about the diffuse emission significantly improves the parameter estimation of the total number of sources $N$ by narrowing the 2$\sigma$ credible interval to $N\sim\left[90,370\right]$ while we find $N\sim\left[66,410\right]$ without diffuse flux information. We notice a similar striking impact on the precision of the detection threshold's posterior, which is much more peaked around the injected value. The quality of the inference on the MSP population's luminosity function remains approximately the same. From a physics perspective, these findings follow from the fact that the assumed diffuse flux is (at least partially) constituted by the sub-threshold population, which we can access this way.

\subsubsection{Impact of incomplete flux measurements}
\label{sec:swyft_pfluxless}

\begin{figure*}[t]
    \begin{centering}
    \includegraphics[width=\linewidth]{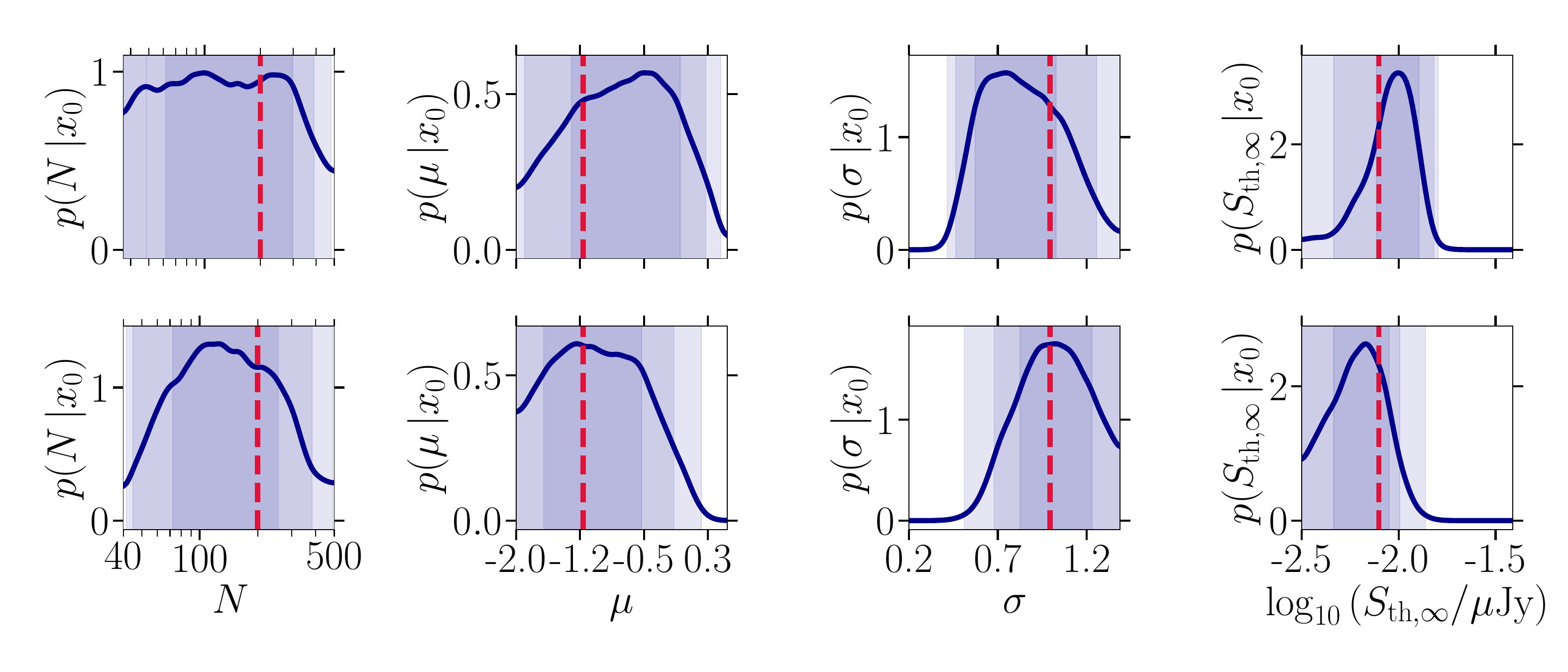}
    \par\end{centering}
    \caption{Impact of incomplete MSP flux measurements. Complete set of one-dimensional marginal posterior distributions adopting the color code of Figure~\ref{fig:mock-swyft-bayesian-comparison} characterizing the radio MSP population model outlined in Section \ref{sec:swyft_simulator} characterized by $N = 200$, $\mu = -1.2$, $\sigma = 1.0$ and $S_{\mathrm{th},\infty} = 9$ $\mu$Jy. We contrast two cases: (\emph{upper row}) $N_\mathrm{detected} = 36$ and $p_{\mathrm{fluxless}} = 0\%$ representing observational results with fewer MSPs than expected regarding the mean expectation of $\langle N_\mathrm{detected}\rangle = 40$. (\emph{lower row}) $N_\mathrm{detected} = 40$ and $p_{\mathrm{fluxless}} = 10\%$ illustrating a scenario where $N_\mathrm{detected} = \langle N_\mathrm{detected}\rangle$ but 10\% of the MSPs lack a flux measurement.
    \label{fig:singleGC-fluxless}}
\end{figure*}

We investigate the impact of increasing $p_{\mathrm{fluxless}}$ as follows: Missing flux information for some detected pulsars implies that the observational source catalog contains fewer sources with complete data than what is expected on average for a given flux threshold $S_{\mathrm{th}, \infty}$. We see this situation in the case of Ter 5. Therefore, we compare in Figure \ref{fig:singleGC-fluxless} \textit{(i)} the inferred one-dimensional marginal posterior distributions for a complete catalog of our baseline mock target but assuming fewer detected sources than possible, $N_\mathrm{detected} = 36$ (upper row) and \textit{(ii)} a case with $N_\mathrm{detected} = 40$ (lower panel) as expected an average but including four sources, i.e.~$p_{\mathrm{fluxless}} = 10\%$, without flux measurement. The corresponding coverage results are shown in Figure \ref{fig:coverage-baseline-pfluxess} of Appendix \ref{app:coverage-plots}, and the median and the 95\% interval credible of $N$ are reported in the fifth and sixth lines of Table \ref{tab:nmed_swyft}.

We observe that the quality of the inference overall improves when including detected sources without derived fluxes. While the total number of sources $N$ is rather unconstrained in the upper panel with $p_{\mathrm{fluxless}} = 0\%$, we find a narrower prior in the lower panel. The width of the 1 and 2$\sigma$ credible intervals are not very different from Figure \ref{fig:singleGC-pflux0p0_wodiff} where we assume a complete catalog of 40 detected sources. This can be understood from the setup of our training data since we generate a data column that counts detected sources no matter the availability of a flux measurement. A similar improvement is found for the two parameters of the MSP population's luminosity function, which are slightly narrower and more centered on the injected value in comparison with the upper row. However, the inferred posterior of the detection threshold deteriorates to some degree as it appears to be wider for the case of $p_{\mathrm{fluxless}} = 0\%$, but the difference is only marginal and, again, comparable to the results shown earlier with full flux information. Therefore, we conclude that adding detected sources irrespective of the availability of an associated flux measurement is beneficial for the ultimate inference.

\subsubsection{Prospects for deeper surveys}
\label{sec:swyft_mock_thresh}

\begin{figure*}[t]
\begin{centering}
\includegraphics[width=0.89\linewidth]{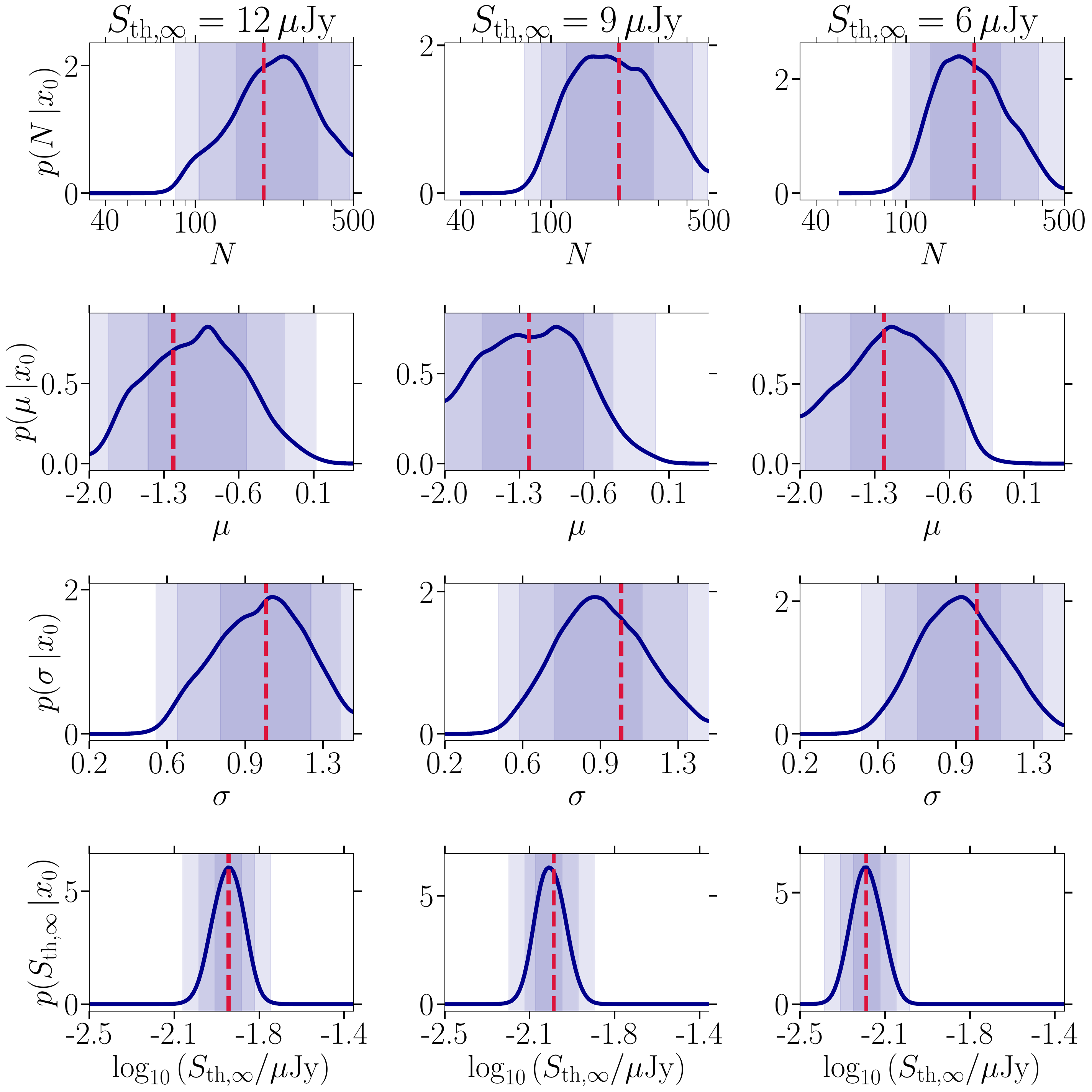}
\par\end{centering}
\caption{Comparison of the complete set of one-dimensional marginal posterior distributions (blue curves) for the four parameters characterizing the radio MSP population model outlined in Section \ref{sec:swyft_simulator}. In each panel we show varying minimal detection threshold for large pulsation periods $S_{\mathrm{th},\infty}$ and accordingly adjusting the expected number of detected sources given the total number of MSPs. The stated values for $N_\mathrm{detected}$ represent the mean expectation for the respective detection threshold obtained from 1000 realizations. The inference has been performed on target mock observations with shared parameters: $p_{\mathrm{fluxless}} = 10\%$, $N = 200$, $\mu = -1.2$, $\sigma = 1.0$ and $S_{\mathrm{diff}} = 2$   mJy. The color code is as in Figure \protect\ref{fig:singleGC-pflux0p0_wodiff}. \textit{(Left column):} $N_\mathrm{detected} = 34$, $S_{\mathrm{th},\infty} = 12$ $\mu$Jy, \textit{(middle column):} $N_\mathrm{detected} = 40$, $S_{\mathrm{th},\infty} = 9$ $\mu$Jy, \textit{(right column):} $N_\mathrm{detected} = 51$, $S_{\mathrm{th},\infty} = 6$ $\mu$Jy.
\label{fig:singleGC-Sthr}}
\end{figure*}

Future surveys will attain lower detection thresholds, which will help us to better determine the MSP populations in GCs. To understand how $S_{\mathrm{th},\infty}$ affects the parameter estimation, we assume a survey that measures the diffuse flux in the GC (again $S_{\mathrm{diff}} = 2$ mJy in a scenario where still $10\%$ of the MSPs are reported without measured flux values). We consider three detection thresholds $S_{\mathrm{th},\infty} = 12$\, $\mu$Jy (left column), $S_{\mathrm{th},\infty} = 9$\, $\mu$Jy (middle column) and $S_{\mathrm{th},\infty} = 6$\, $\mu$Jy (right column) as shown in terms of one-dimensional marginal posteriors in Figure~\ref{fig:singleGC-Sthr} (the corresponding coverage results are shown in Figure \ref{fig:coverage-baseline-threshold} of Appendix \ref{app:coverage-plots}), and the median and the 95\% credible of $N$ are reported in lines seven to nine of Table \ref{tab:nmed_swyft}. We show results for $S_{\mathrm{th},\infty} = 12$\, $\mu$Jy -- which is worse than the assumed value of our baseline mock setup -- for pedagogical reasons to better illustrate the evolution of the posteriors with decreasing detection threshold. Since a given detection threshold uniquely determines the mean number of detected MSPs, we adjust the assumed value of $N_\mathrm{detected}$ as described in Section \ref{sec:swyft-initial-mock-data}. Consequently, we find $N_\mathrm{detected} = 34, 40$ and 51 for the three considered values of $S_{\mathrm{th},\infty}$, respectively.

The most striking improvement is achieved when lowering the detection threshold from 9 $\mu$Jy to 6 $\mu$Jy while the initial step from 12 $\mu$Jy to 9 $\mu$Jy does not have sizeable effects on the final posteriors. In fact, the second step to 6 $\mu$Jy adds 11 MSPs to the source catalog while the first step only adds 6. We find that the posterior distribution of $N$ becomes much narrower positioning the $2\sigma$ credible interval between 100 and 390 MSPs. Improving the detection threshold by a factor of two renders this parameter accessible in the posterior estimation with a defined 68\% and 90\% credibility interval. This is a natural consequence of slightly increasing the number of detected sources because it increases the part of the luminosity function accessible to the inference pipeline. Note that this setup uses 10\% of fluxless MSPs as well as the inclusion of a measurement of the diffuse flux for the mock GC. The inference on the remaining three parameters becomes better but to a lesser extent.

\subsection{Real data: Terzan 5}
\label{sec:results-swyft-Ter5}

\begin{figure*}[t]
\begin{centering}
\includegraphics[width=0.89\linewidth]{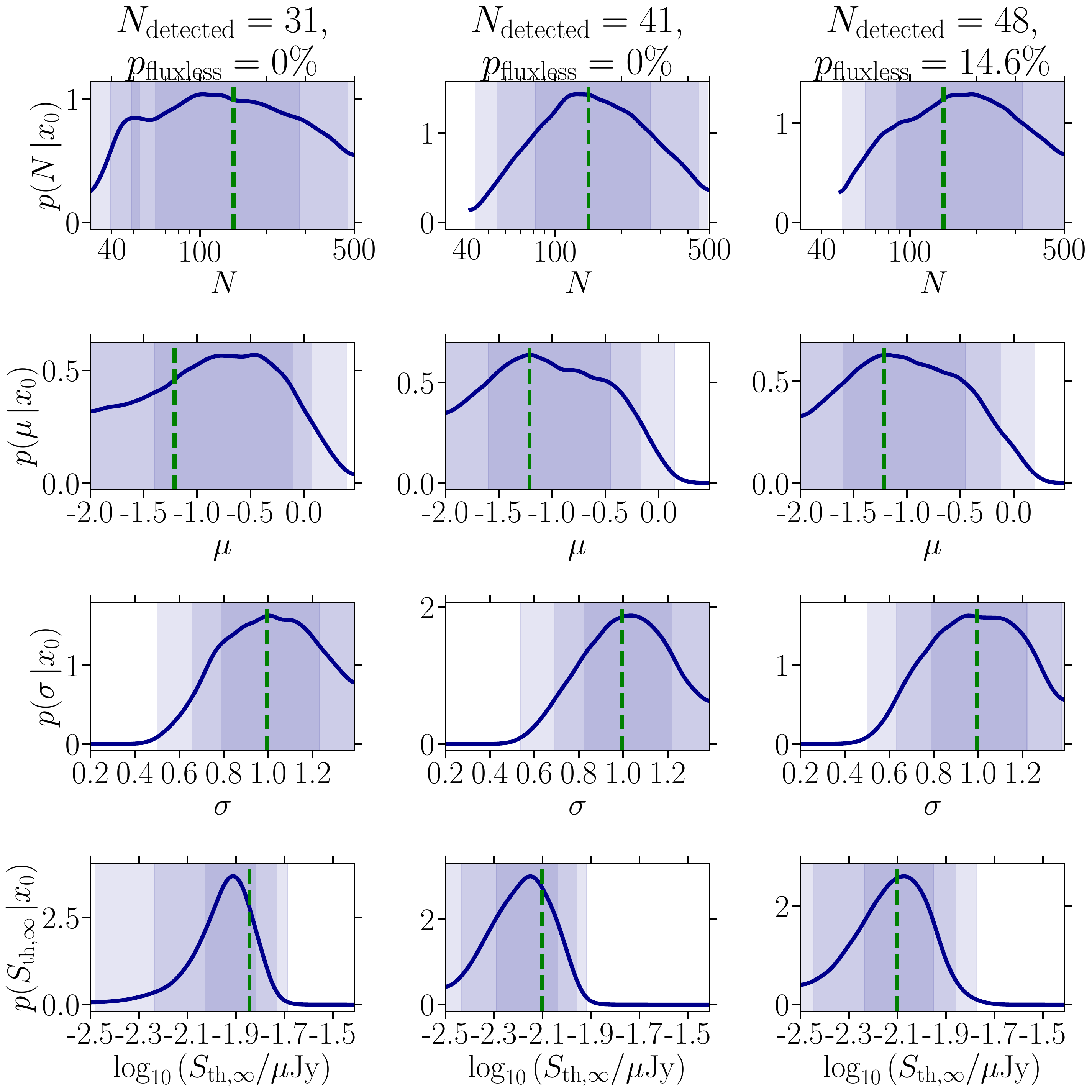}
\par\end{centering}
\caption{Comparison of the complete set of one-dimensional marginal posterior distributions (blue curves) for the four parameters characterizing the radio MSP population model outlined in Section \ref{sec:swyft_simulator} obtained for real MSP catalogs of Terzan 5 outlined in Section \ref{sec:ter5-real-sample}. We consider three datasets: (\textit{left column}:) Original sample of 31 MSPs with flux measurements~\citep{2022ApJ...941...22M} re-scaled to a frequency of 1284 MHz, (\textit{middle column}:) 41 MSPs with flux measurements encompassing the~\cite{2022ApJ...941...22M} sample and 10 additional sources characterized by MeerKAT at the same frequency and (\textit{right column}:) Adding 7 additional sources to the sample without flux measurement ($p_{\mathrm{fluxless}} \approx 14.6\%$) to account for all sources currently known MSPs in Terzan 5. The dashed, vertical, green lines for $N$, $\mu$ and $\sigma$ denote the inference results of \cite{2013MNRAS.431..874C} while the respective line for $S_{\mathrm{th},\infty}$ is the one we used. The inference has been performed on a target mock observation without a diffuse measurement. The color code is as in Figure \protect\ref{fig:singleGC-pflux0p0_wodiff}.
}
\label{fig:swyft-terzan5-results}
\end{figure*}

Having tested the performance of the \sbi~implementation on purely simulated data, we now venture to apply the approach to the available data for Terzan 5 presented in Section \ref{sec:ter5-real-sample}. To show the evolution of the posterior distributions with the amount of available data, {\it i.e.}, an increasing number of detected sources with and without flux measurement, we define three datasets: \textit{(i)} The 31 MSPs originally listed by~\cite{2022ApJ...941...22M}; \textit{(ii)} adding the 10 MSPs analyzed by MeerKAT and \textit{(iii)} using all 48 known MSPs in Terzan 5; seven of them without flux measurements -- $p_{\mathrm{fluxless}}\approx 14.6\%$. To ensure comparability to previous results on mock data, we keep the prior distributions outlined in Table \ref{tab:swyft_priors}. In particular, this implies $N\in\left[N_{\rm detected}, 500\right]$ and $\mu\in\left[-2.0, 0.5\right]$. This also allows us to easily compare the \sbi~results with previous work in \cite{2013MNRAS.431..874C} and the application of our Bayesian approach in Section \ref{sec:bayes_res}. The results are displayed in Figure \ref{fig:swyft-terzan5-results} and the median and the 95\% credible intervals of $N$, $\mu$ and $\sigma$ are reported in Table \ref{tab:nmed_real_swyft}. The corresponding coverage results are shown in Figure \ref{fig:coverage-ter5} of Appendix \ref{app:coverage-plots}.

To facilitate the comparison with the results obtained by the authors of \cite{2013MNRAS.431..874C}, we display their inferred median parameters for Terzan 5 as dashed vertical green lines. As concerns the posterior distribution for the total number of MSPs in Terzan 5, we observe that increasing the sample leads to a better definition of the posterior distribution, which evolves from a rather broad posterior in case \textit{(i)} to a well-defined two-sided posterior in cases \textit{(ii)} and \textit{(iii)}. This behavior follows our inference findings on the baseline mock MSP population. In data selection \textit{(iii)}, we obtain a $2\sigma$ credible interval of $N\in\left[54, 452\right]$.  We want to emphasize here that with the stated Ter 5 catalogs we are in a different situation than with our mock datasets. It is not clear that the number of detected sources associated with the assumed value of $S_{\mathrm{th}, \infty}$ (based on the lowest flux value in the list of detected MSPs) is indeed the average number of detected sources one would expect for the true population parameters of Ter 5. It is much more likely that we are always below the average expectations and much closer to the scenario considered in the upper panel of Figure \ref{fig:singleGC-fluxless}. The quality of our inferred posteriors might suffer from a certain out-of-domain effect regarding simulated training samples and reality. Yet the coverage of our \sbi~pipeline is well-calibrated (see Appendix \ref{app:coverage-plots}) so that the results presented here are the best we are currently able to derive.

With respect to the sizable improvement from \emph{(i)} to \emph{(ii)}, the evolution of the posteriors for the parameters associated with the luminosity function shows only a mild gain when going from \emph{(ii)} to \emph{(iii)}. Our posteriors are consistent with the results of \cite{2013MNRAS.431..874C}. In this case, the inclusion of fluxless MSPs is not as beneficial as having complete flux information because it renders the inference on the detection threshold less constraining. This effect we could also observe in our study of the impact of $p_{\mathrm{fluxless}}$ in Figure \ref{fig:singleGC-fluxless} where the lower row shows wider posterior for the detection threshold than the corresponding upper row. We thus argue that obtaining a complete assessment of radio fluxes is essential to pinpoint the properties of the MSP population in Terzan 5 with \sbi. 

\begin{table}[t]
    \centering
    \begin{tabular}{lccc}
        \Xhline{5\arrayrulewidth}
        Case & $N$ &  $\mu$ & $\sigma$\\
        $N_{\mathrm{detected}}=31$ & 126$^{+320}_{-89}$ & -0.81$^{+1.01}_{-1.12}$ & 1.01$^{+0.35}_{-0.40}$\\
        $N_{\mathrm{detected}}=41$ & 146$^{+283}_{-94}$ & -1.05$^{+0.99}_{-0.89}$ & 1.02$^{+0.34}_{-0.38}$\\
        $N_{\mathrm{detected}}=48$ & 158$^{+294}_{-104}$ & -1.02$^{+1.02}_{-0.91}$ & 1.00$^{+0.36}_{-0.39}$\\
        \hline

    \end{tabular}
    \caption{Median and 95\% credible interval of the number of MSPs $N$, the mean $\mu$ and the width $\sigma$ of the luminosity function obtained through the \sbi~analysis applied to real data for various cases.}
    \label{tab:nmed_real_swyft}
\end{table}

\section{Discussion}
\label{sec:discussion}

The current limiting factor in our understanding of the population of MSPs in GCs is the observational data. The available MSP catalogs are limited by the performance of radio instruments and by their non-uniformity. Unsurprisingly, our ability to robustly infer the number of MSPs hosted by a cluster will increase as the detection threshold of radio telescopes decreases (Section \ref{sec:swyft_mock_thresh}). Not only an improved detection threshold can improve the inference, but also a more systematic measurement of pulsar fluxes. We showed that the fraction of sources not having a flux measurement is a key element in the inference pipeline both on mock and real data, 
Sections~\ref{sec:swyft_pfluxless}, \ref{sec:results-swyft-Ter5}.

Moreover, as we have seen in section \ref{sec:swyft_diffuse}, incorporating a measurement of the diffuse radio emission can be a crucial element of the parameter estimation. Terzan 5 can be found in the NRAO VLA Sky Survey \citep[NVSS,][]{1998AJ....115.1693C} catalog as NVSS 174804-244641, a source with a total flux density of $S_\mathrm{tot} = 3.4$ mJy at 1.4 GHz. We note that $S_\mathrm{det}$, the sum of all MSP fluxes measured in Terzan 5, is already larger than 3.4 mJy. This apparent inconsistency is probably partly due to the fact that the position and the size of NVSS 174804-244641 do not exactly match those of Terzan 5 commonly assumed today. We thus argue that new measurements of the total radio emission of the GC will highly benefit the analysis of the properties of its MSP population. Nonetheless, one has to keep in mind that, while MSPs seem to be the main contributors to the radio emission of GCs, other populations could contribute, such as low-mass X-ray binaries and accreting stellar mass black holes \citep{2020ApJ...904..147U}. Finally, data of several GCs could be combined, if the luminosity function is assumed to be unique, as done by \cite{2011MNRAS.418..477B}. This will enable the usage of an extended dataset, if uniform and accessible. If the detection threshold at large periods $S_{\mathrm{th},\infty}$ does not vary too much from one GC to the next, the luminosity of the dimmest object in each cluster can fluctuate as their distance varies. Therefore, each GC probes a different region of the luminosity function. In the Bayesian analysis framework, each cluster has its own independent likelihood $\mathcal{L}_j$, and the total likelihood is the product of all these likelihoods (see Appendix \ref{app:multiple_gc} for more details). With \sbi, only minor modifications of the simulator will instead be necessary.

After the publication of \cite{2013MNRAS.431..874C}, stellar data from the \textit{Gaia} mission made it possible to derive a much more precise distance measurement for Ter 5. While we kept for the sake of comparability the distance of $(5.5\pm0.9)$ kpc throughout our Bayesian and \sbi~applications, a better distance characterization is given by $d=(6.620\pm0.150)$ kpc in \cite{2021MNRAS.505.5957B}. These two measurements are consistent, but the more recent one puts Ter 5 about 1 kpc farther than we assume in our analyses. Yet, Eq.~\ref{eq:logl_to_logs} states that a different distance translates to a mere shift of an MSP's measured flux $\log{S}$. Therefore, we do not expect strong qualitative differences in the inference results except for the mean $\mu$ of the MSP luminosity function. To quantitatively probe the impact of this updated distance to Ter 5, we inferred the posteriors for the case of $N_{\rm detected} = 32$ within this setting. The results are shown in Figure \ref{fig:ter5-new-distance} (plus coverage) of Appendix \ref{app:ter5-new-distance}. As expected, the inference results are slightly different from the ones visualized in the corresponding plot in Figure \ref{fig:swyft-terzan5-results} except for the results on $\mu$. Here, the posterior's maximum is shifted to larger fluxes, which is expected due to the increased distance of Ter 5 compared to Figure \ref{fig:swyft-terzan5-results}. We conclude that our results are robust against a re-definition of the distance to a single GC and all plots shown in the main text retain their validity.

An improvement of our work towards more realism lies in our implementation of the detection threshold. While \cite{2013MNRAS.431..874C} chose to work with an MSP-\textit{independent} threshold, we have implemented an MSP-\textit{dependent} threshold to mimic the radiometer equation. However, the variations from one threshold to the next are random and do not depend on specific parameters of the MSP. As noted above, we know that these thresholds strongly depend on the pulsation period of pulsars. Moreover, the luminosity of radio pulsars should also depend on their period and period derivative \citep[see \textit{e.g.}][]{2013IJMPD..2230021B}. Therefore, further studies could benefit from and include a simulation of the pulsation period of MSPs and its link with the radio emission. Theoretical works and simulations could help in that respect to guarantee that the inference is robust and correct.

\section{Conclusions}
\label{sec:conclusions}

The question of the number of MSPs in GCs has been studied by different groups via their luminosity function, each method producing different (but not incompatible) results. In this work, we tackled this problem through an analysis framework which quickly developed in the last decade: likelihood-free/simulation-based inference. Our method produces Bayesian posteriors which quantify the uncertainty on various parameters of interest. While the method used by \cite{2011MNRAS.418..477B} could be interpreted as part of SBI, it does not produce Bayesian posteriors. As for the work of \cite{2013MNRAS.431..874C}, based on a Bayesian likelihood, we have shown in Section \ref{sec:bayes_res} that it likely produces posteriors that are too conservative indicating a poor calibration of the statistical analysis. 
With therefore developed a novel analysis pipeline based on \sbi. We tested our method on simulated datasets and we demonstrated that our analysis consistently produces well-behaved posteriors. In the mock datasets analysis, the best improvements are achieved when the simulated datasets include a measurement of the diffuse radio emission of the GC, or when we assume a detection threshold at large periods of $\sim 6$ $\mu$Jy, that is, twice as low as current surveys. 
Applied to real data of Terzan 5, our analysis robustly hints at a population of about 158 MSPs. Despite the well-behaved posteriors, however, large uncertainties remain in the determination of the 95\% credible intervals of the number of MSPs.
On the other hand, our 95\% credible intervals of the width of the luminosity function remain very similar to those of \cite{2013MNRAS.431..874C}, while the upper limit on the mean of the luminosity function can be better determined.

Our results call for dedicated campaigns of flux measurements of MSPs in GCs, as well as deep imaging observations associated with a measurement of the GC radio, diffuse emission.

\section*{Acknowledgments}

We acknowledge useful discussion with N.~Anau Montel on the \sbi~implementation technical aspects, as well as with K. Dage on the population of stars in Terzan 5. The authors acknowledge support from the Embassy of France in Canada via the 2022 New Research Collaboration Program from the France Canada Research Fund. JB and CE acknowledge financial support from the Programme National des Hautes Energies of CNRS/INSU with INP and IN2P3, co-funded by CEA and CNES, from the ‘Agence Nationale de la Recherche’, grant number ANR-19-CE310005-01 (PI: F.~Calore), and the Centre National d’Etudes Spatiales (CNES). The work of JB is supported by NASA under award number 80GSFC21M0002. The work of CE has been supported by the EOSC Future project which is co-funded by the European Union Horizon Programme call INFRAEOSC-03-2020, Grant Agreement 101017536. This publication is supported by the European Union's Horizon Europe research and innovation programme under the Marie Sk\l odowska-Curie Postdoctoral Fellowship Programme, SMASH co-funded under the grant agreement No.~101081355. MC acknowledges financial support from the Centre National d’Etudes Spatiales (CNES). DH acknowledges funding from the Natural Sciences and Engineering Research Council of Canada (NSERC) and the Canada Research Chairs (CRC) program. This work has been completed thanks to the facilities offered by the Univ.~Savoie Mont Blanc
- CNRS/IN2P3 MUST computing center, especially with GPUs acquired thanks to the Labex Enigmass R\&D Booster program.


\bibliography{GC_swyft}{}
\bibliographystyle{aasjournal}



\appendix

\section{Flux dataset}
\label{app:flux_dataset}

In Table \ref{tab:psr_fluxes}, we provide the data of the pulsars in Terzan 5 used in this work.

\begin{table}[p]
    \centering
    \begin{tabular}{ccccc}
        \Xhline{5\arrayrulewidth}
        Pulsar & S$_{1284}$ & S$_{1400}$ & S$_{2000}$ & $\alpha$ \\
         & ($\mu$Jy) & ($\mu$Jy) & ($\mu$Jy) & \\
        \hline
         A 	&	--	&	2700	&	1700	&	 -1.31(15) 	\\
         C 	&	--	&	1100	&	670	&	 -1.4(2) 	\\
         D 	&	--	&	71	&	45	&	 -1.28(14) 	\\
         E 	&	--	&	170	&	110	&	 -1.16(13) 	\\
         F 	&	--	&	55	&	35	&	 -1.218(95) 	\\
         G 	&	--	&	24	&	22	&	 -0.26(11) 	\\
         H 	&	--	&	39	&	24	&	 -1.33(9) 	\\
         I 	&	--	&	95	&	55	&	 -1.53(11) 	\\
         K 	&	--	&	66	&	39	&	 -1.47(7) 	\\
         L 	&	--	&	96	&	43	&	 -2.26(7) 	\\
         M 	&	--	&	140	&	91	&	 -1.14(11) 	\\
         N 	&	--	&	150	&	100	&	 -1.02(11) 	\\
         O 	&	--	&	310	&	160	&	 -1.9(2) 	\\
         Q 	&	--	&	56	&	36	&	 -1.24(14) 	\\
         R 	&	--	&	35	&	17	&	 -2.07(14) 	\\
         S 	&	--	&	20	&	14	&	 -1.09(13) 	\\
         T 	&	--	&	26	&	15	&	 -1.61(9) 	\\
         U 	&	--	&	30	&	12	&	 -2.491(97) 	\\
         V 	&	--	&	100	&	77	&	 -0.86(7) 	\\
         W 	&	--	&	54	&	31	&	 -1.53(14) 	\\
         X 	&	--	&	43	&	24	&	 -1.63(7) 	\\
         Y 	&	--	&	37	&	29	&	 -0.76(12) 	\\
         Z 	&	--	&	30	&	23	&	 -0.7(2) 	\\
         aa 	&	--	&	29	&	20	&	 -1.00(14) 	\\
         ab 	&	--	&	45	&	23	&	 -1.83(12) 	\\
         ac 	&	--	&	31	&	17	&	 -1.67(15) 	\\
         ae 	&	--	&	56	&	50	&	 -0.3(2) 	\\
         af 	&	--	&	33	&	22	&	 -1.19(12) 	\\
         ag 	&	--	&	16	&	9.2	&	 -1.6(2) 	\\
         ah 	&	--	&	14	&	7.1	&	 -1.9(2) 	\\
         ai 	&	--	&	33	&	28	&	 -0.4(2) 	\\
         \hline
        ao	&	12	&	--	&	--	&	--	\\
        ap	&	15	&	--	&	--	&	--	\\
        au	&	12	&	--	&	--	&	--	\\
        ax	&	8	&	--	&	--	&	--	\\
        aq	&	17	&	--	&	--	&	--	\\
        ar	&	44	&	--	&	--	&	--	\\
        at	&	19	&	--	&	--	&	--	\\
        as	&	10	&	--	&	--	&	--	\\
        av	&	15	&	--	&	--	&	--	\\
        aw	&	10	&	--	&	--	&	--	\\

        \hline

    \end{tabular}
    \caption{Flux densities of Terzan 5 pulsars used in our \sbi~analysis. All data from the top section are from \cite{2022ApJ...941...22M} and all data from the bottom section are from \cite{2024arXiv240317799P}.}
    \label{tab:psr_fluxes}
\end{table}

\section{Collection of coverage results for all scenarios}
\label{app:coverage-plots}

In this appendix, we collect and provide coverage results for all radio MSP population scenarios we tested and performed inference on. The parameters of the MSP scenario are stated in the respective plot's caption.

\begin{figure}[h]
    \centering\includegraphics[width=0.49\columnwidth]{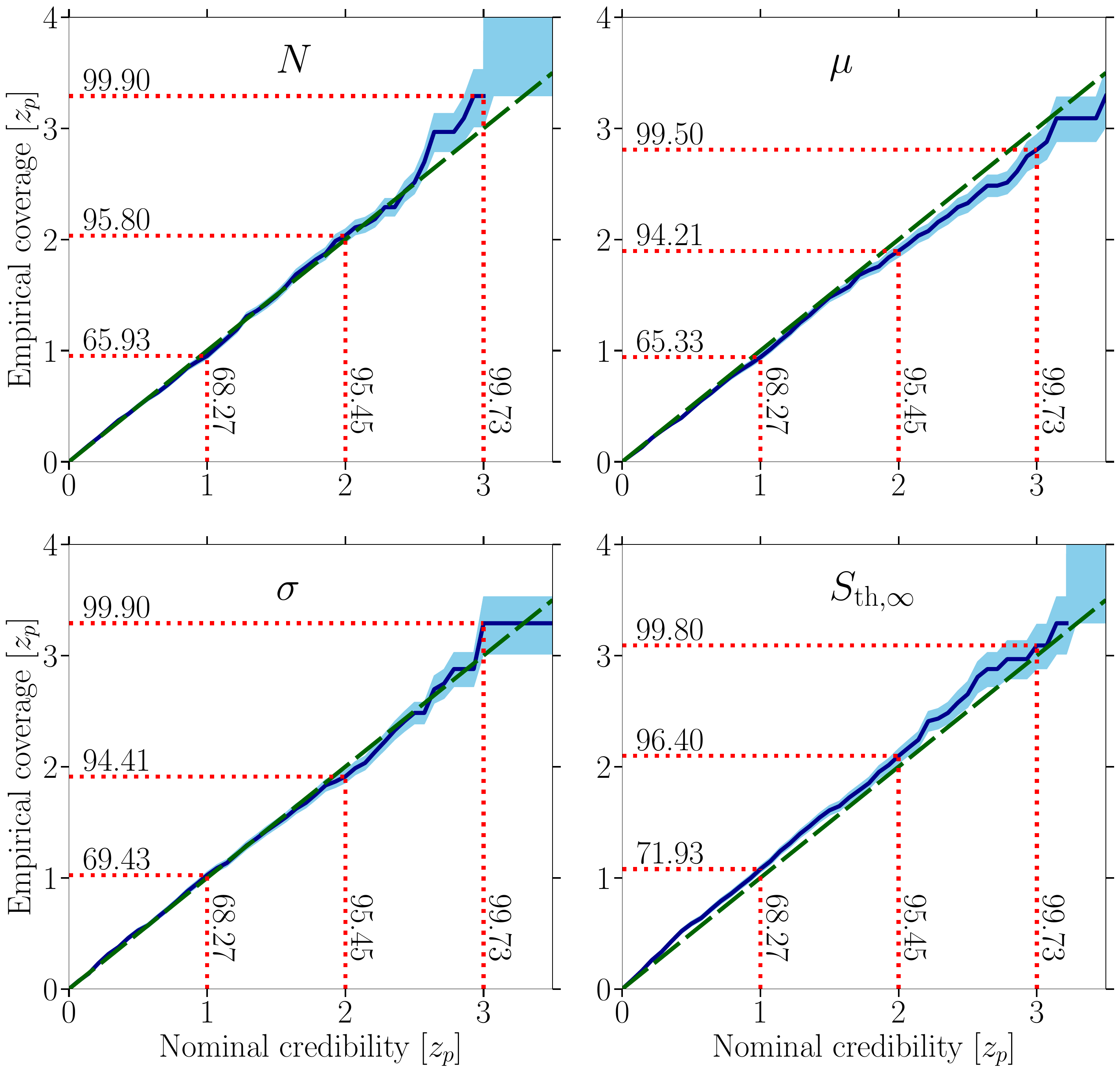}
    \caption{Same as the right panel of Figure \ref{fig:mock-swyft-bayesian-comparison} in the main text for the baseline mock data setup characterized by: $N_\mathrm{detected} = 40$, $p_{\mathrm{fluxless}} = 0\%$, $N = 200$, $\mu = -1.2$, $\sigma = 1.0$ and $S_{\mathrm{th},\infty} = 9$ $\mu$Jy with a prior range $N\in\left[N_\mathrm{detected}, 500\right]$ (see Figure \ref{fig:singleGC-pflux0p0_wodiff} for the parameter inference).}
    \label{fig:coverage-baseline_woDiff}
\end{figure}

\begin{figure}
    \centering
    \includegraphics[width=0.49\columnwidth]{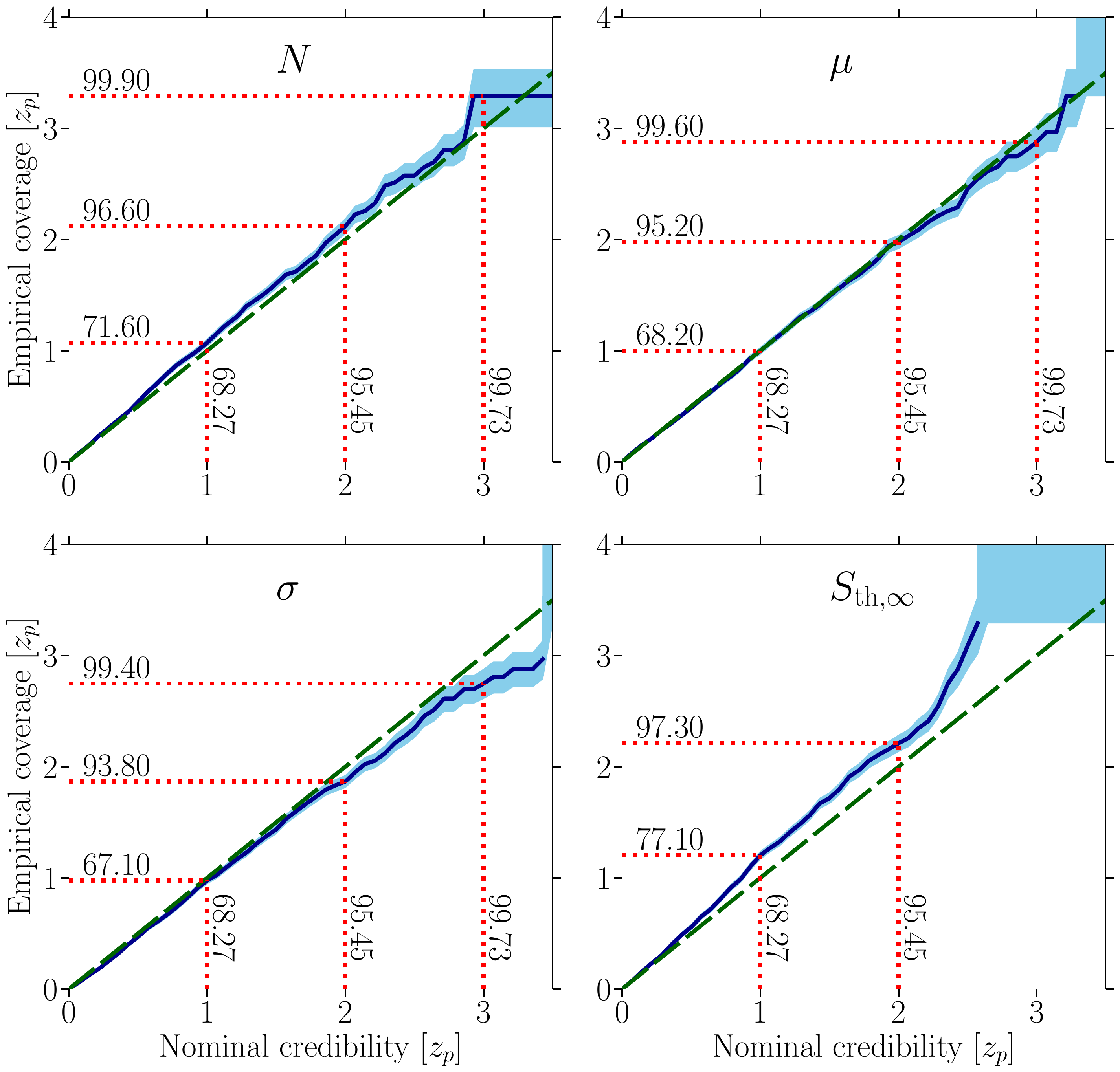}
    \caption{Same as the right panel of Figure \ref{fig:mock-swyft-bayesian-comparison} in the main text for the baseline mock data setups exploring the impact of adding a diffuse measurement for the GC under scrutiny with results shown in Figure \ref{fig:singleGC-pflux0p0_wdiff}: $N_\mathrm{detected} = 40$, $p_{\mathrm{fluxless}} = 0\%$, $N = 200$, $\mu = -1.2$, $\sigma = 1.0$ and $S_{\mathrm{th},\infty} = 9$ $\mu$Jy and $S_{\mathrm{diff}} = 2$ mJy.}
    \label{fig:coverage-baseline-wDiff}
\end{figure}

\begin{figure*}
    \centering
    \includegraphics[width=0.49\linewidth]{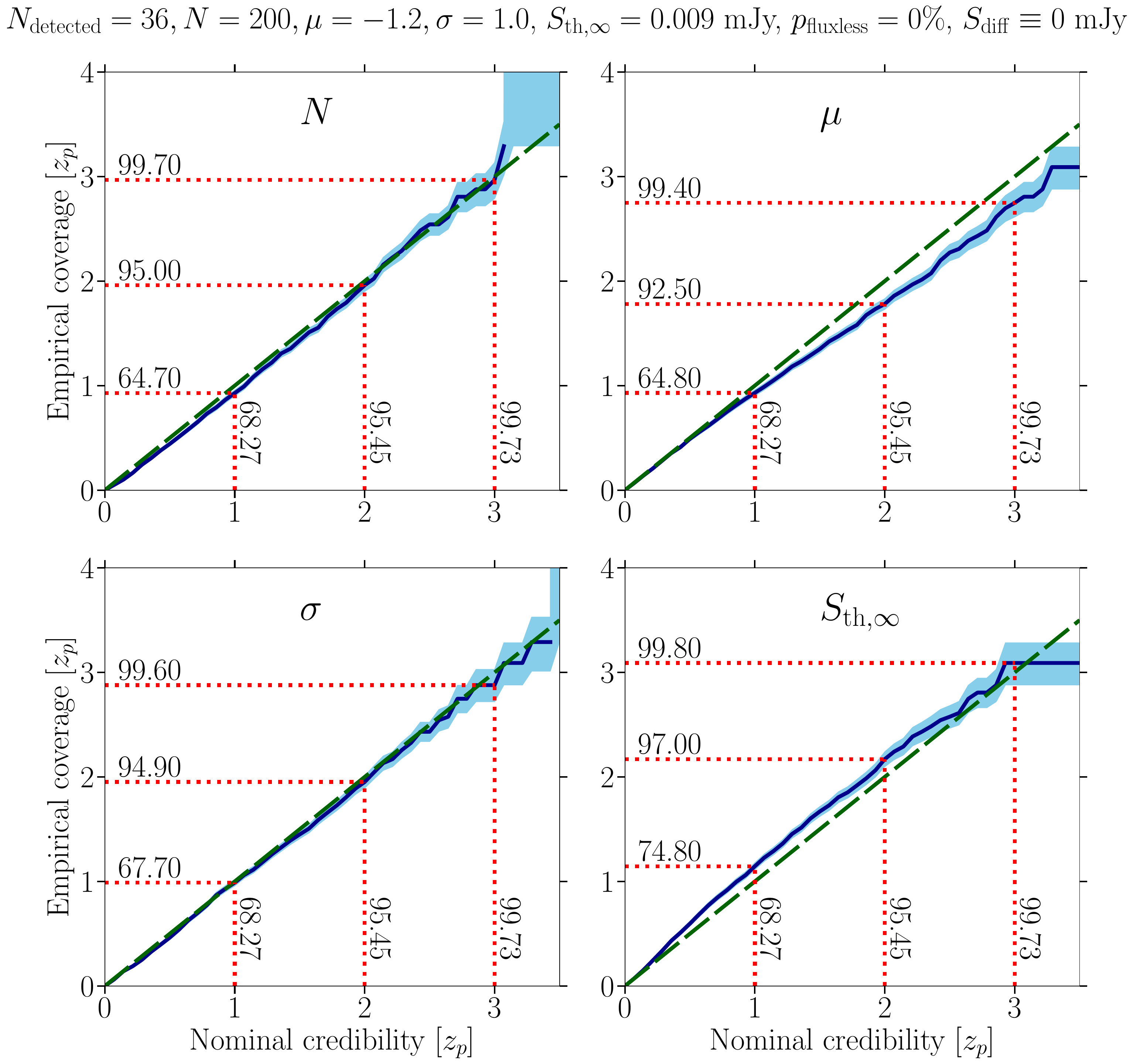}\hfill\includegraphics[width=0.49\linewidth]{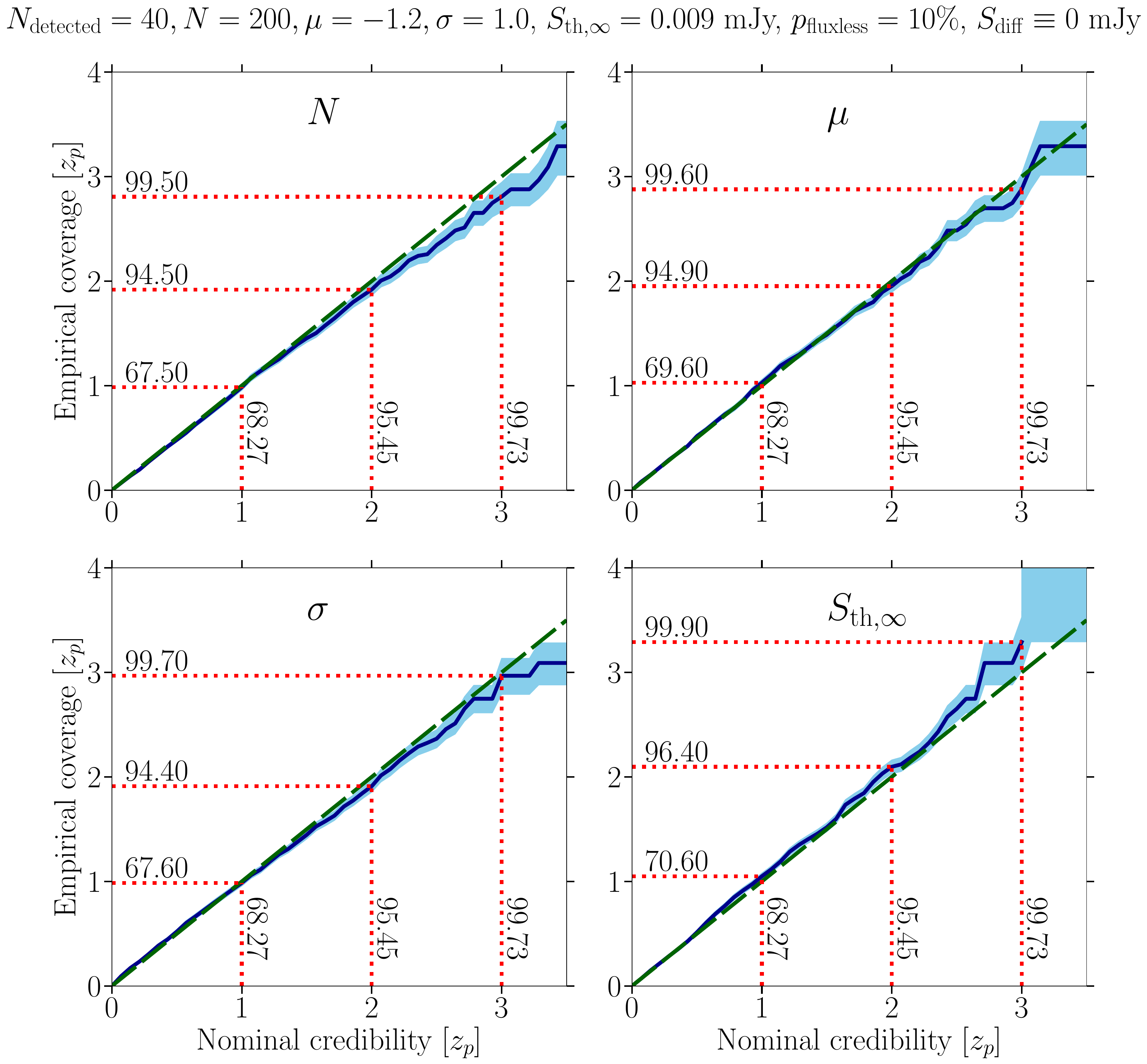}
    \caption{Same as the right panel of Figure \ref{fig:mock-swyft-bayesian-comparison} in the main text for the baseline mock data setups exploring the impact of having an MSP catalog lacking flux information with results shown in Figure \ref{fig:singleGC-fluxless}: (\emph{left}:) $N_\mathrm{detected} = 36$, $p_{\mathrm{fluxless}} = 0\%$, $N = 200$, $\mu = -1.2$, $\sigma = 1.0$ and $S_{\mathrm{th},\infty} = 9$ $\mu$Jy; (\emph{right}:) $N_\mathrm{detected} = 40$, $p_{\mathrm{fluxless}} = 10\%$, $N = 200$, $\mu = -1.2$, $\sigma = 1.0$ and $S_{\mathrm{th},\infty} = 9$ $\mu$Jy.}
    \label{fig:coverage-baseline-pfluxess}
\end{figure*}

\begin{figure*}
    \centering
    \includegraphics[width=0.49\linewidth]{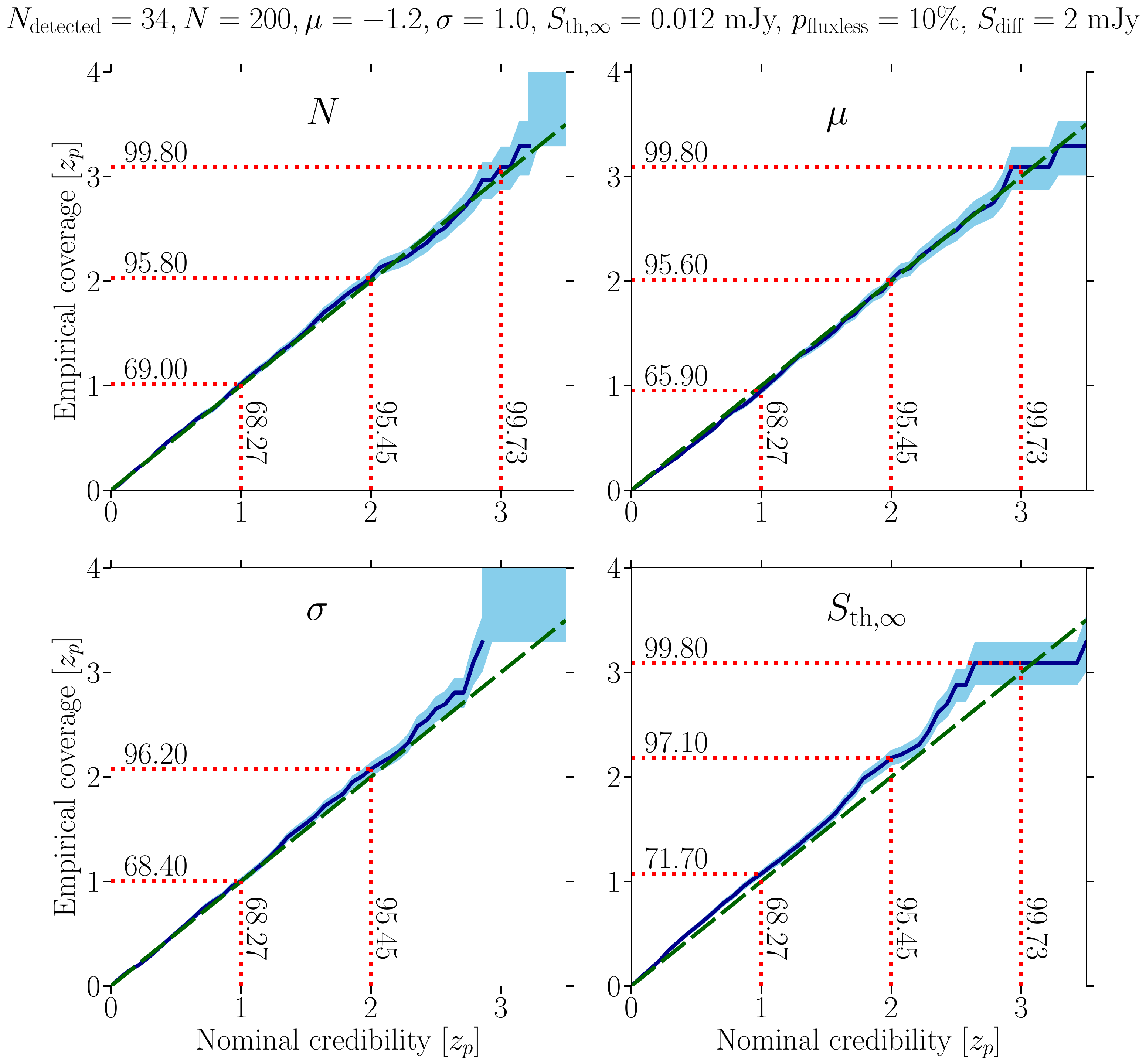}\hfill\includegraphics[width=0.49\linewidth]{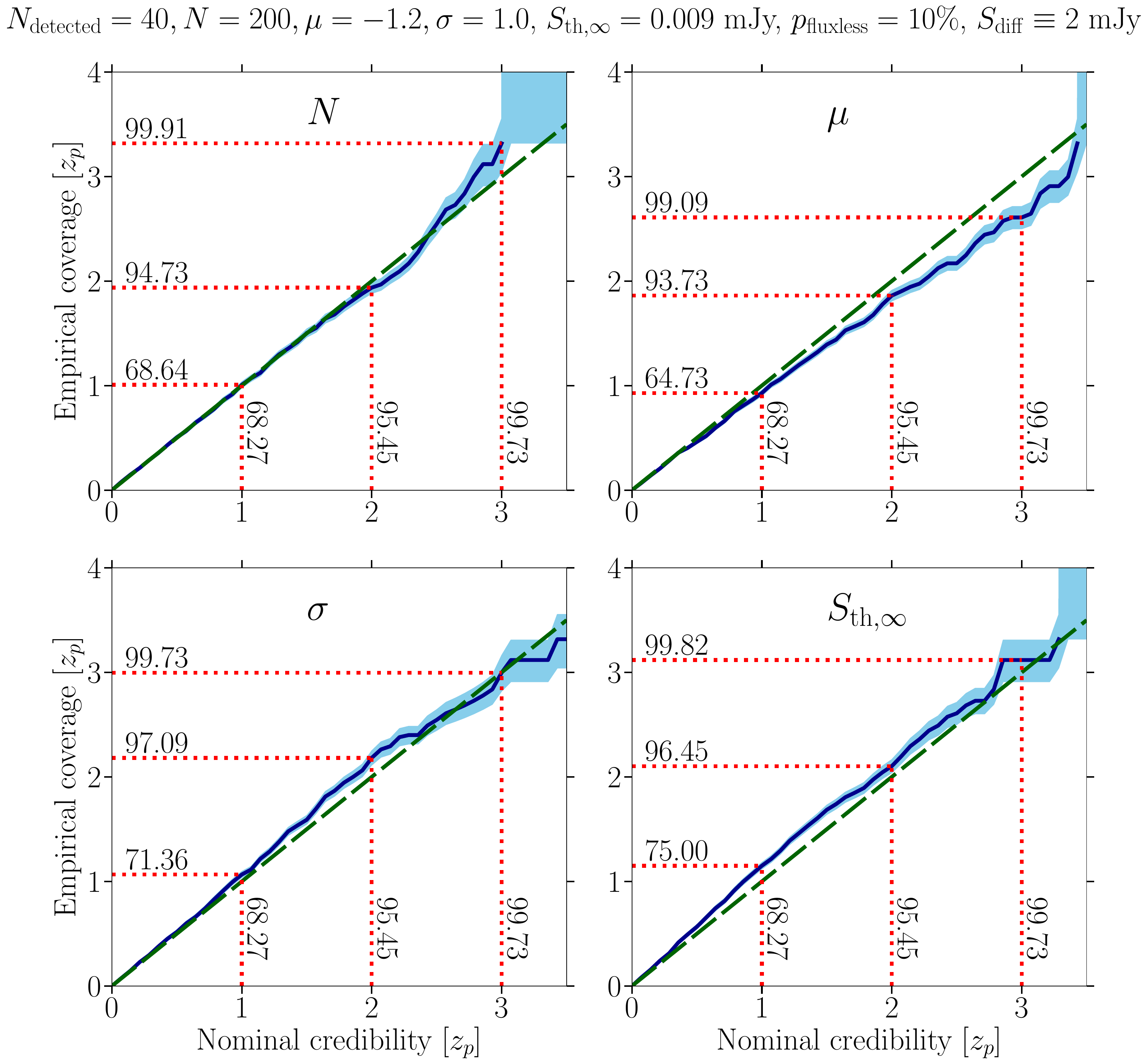}\\
    \includegraphics[width=0.49\linewidth]{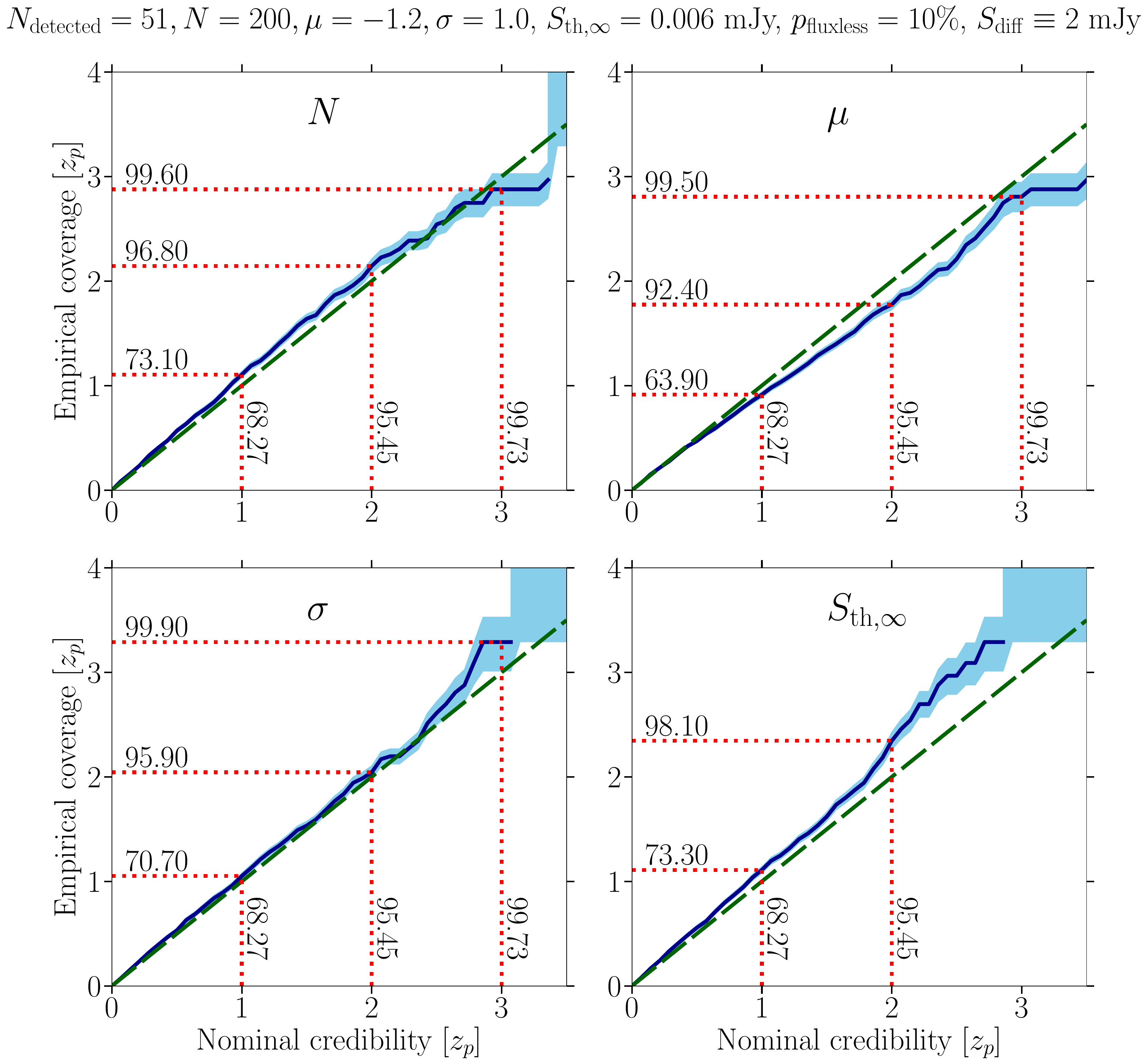}
    \caption{Same as the right panel of Figure \ref{fig:mock-swyft-bayesian-comparison} in the main text for the baseline mock data setups exploring the impact of lowering the detection threshold with results shown in Figure \ref{fig:singleGC-Sthr}: (\emph{upper left}:) $N_\mathrm{detected} = 34$, $p_{\mathrm{fluxless}} = 10\%$, $N = 200$, $\mu = -1.2$, $\sigma = 1.0$ and $S_{\mathrm{th},\infty} = 12$ $\mu$Jy and $S_{\mathrm{diff}} = 2$ mJy; (\emph{upper right}:) $N_\mathrm{detected} = 40$, $p_{\mathrm{fluxless}} = 10\%$, $N = 200$, $\mu = -1.2$, $\sigma = 1.0$ and $S_{\mathrm{th},\infty} = 9$ $\mu$Jy and $S_{\mathrm{diff}} = 2$ mJy and (\emph{bottom}:) $N_\mathrm{detected} = 51$, $p_{\mathrm{fluxless}} = 10\%$, $N = 200$, $\mu = -1.2$, $\sigma = 1.0$ and $S_{\mathrm{th},\infty} = 6$ $\mu$Jy, $S_{\mathrm{diff}} = 2$ mJy.}
    \label{fig:coverage-baseline-threshold}
\end{figure*}

\begin{figure*}
    \centering
    \includegraphics[width=0.49\linewidth]{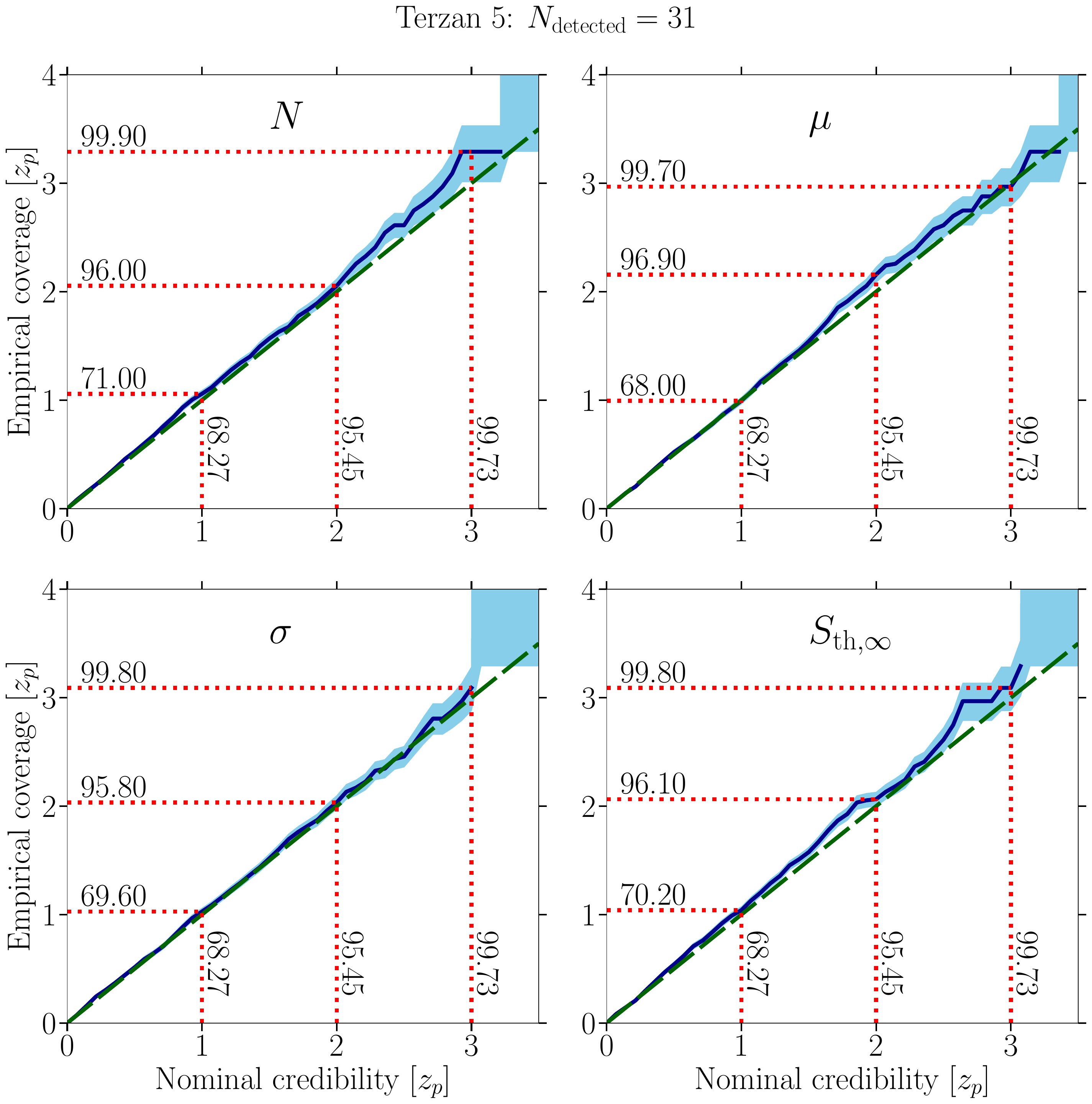}\hfill\includegraphics[width=0.49\linewidth]{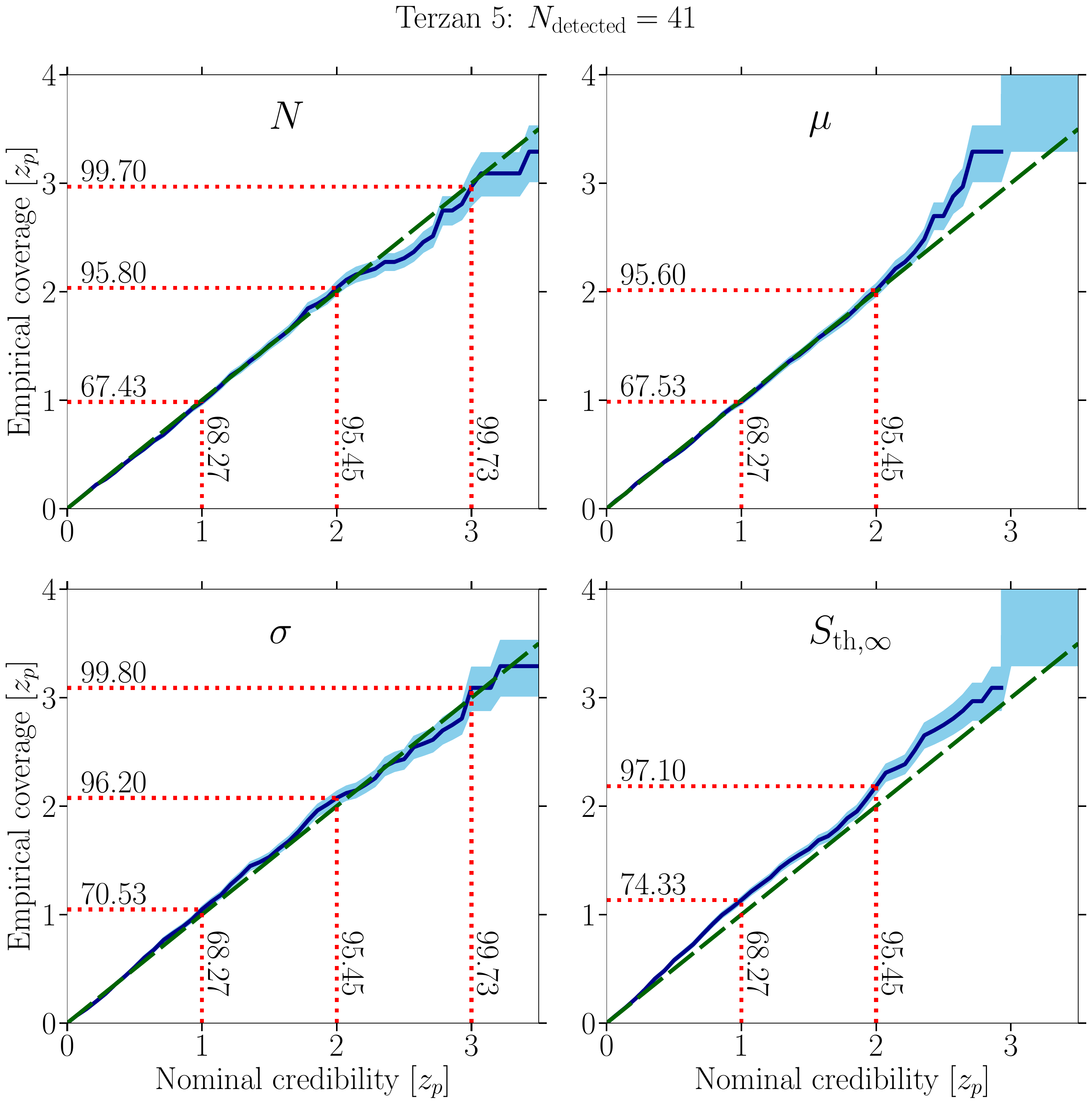}\\
    \includegraphics[width=0.49\linewidth]{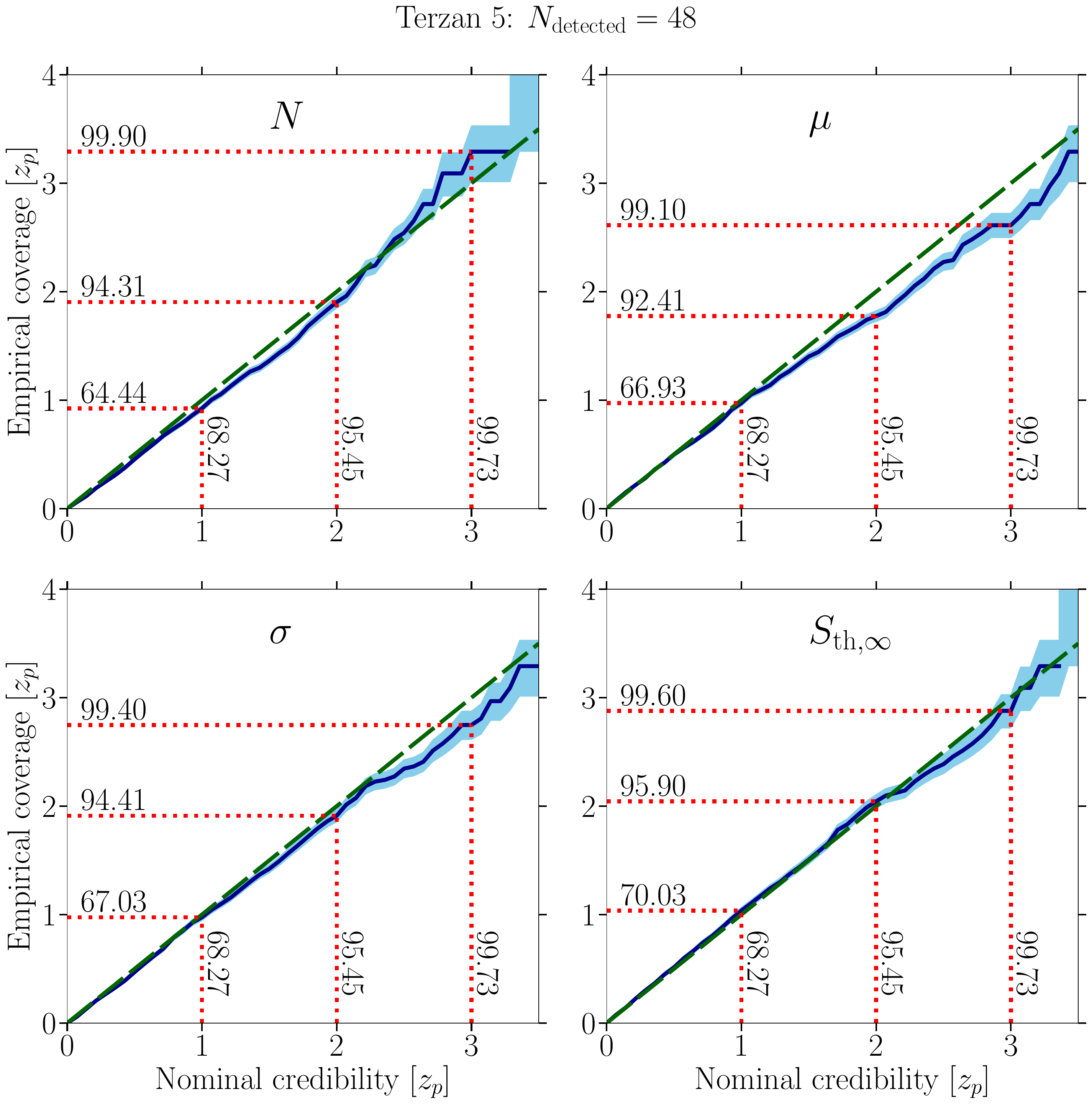}
    \caption{Same as the right panel of Figure \ref{fig:mock-swyft-bayesian-comparison} in the main text for the radio MSP population of Terzan 5 with: (\emph{upper left}:) $N_\mathrm{detected} = 32$, $p_{\mathrm{fluxless}} = 0\%$ and $S_{\mathrm{th},\infty} = 16.5$ $\mu$Jy; (\emph{upper right}:) $N_\mathrm{detected} = 41$, $p_{\mathrm{fluxless}} = 0\%$ and $S_{\mathrm{th},\infty} = 8$ $\mu$Jy and (\emph{bottom}:) $N_\mathrm{detected} = 48$, $p_{\mathrm{fluxless}} = 14.6\%$ and $S_{\mathrm{th},\infty} = 8$ $\mu$Jy. See Figure \ref{fig:swyft-terzan5-results} for the parameter inference results.}
    \label{fig:coverage-ter5}
\end{figure*}

\section{Enlarging the prior range on the total number of millisecond pulsars}
\label{app:wide-priors}

\begin{figure*}
    \centering
    \includegraphics[width=0.49\columnwidth]{posteriors_targetSIM_model_Ndetected_40_Sthr_0.009_pfluxless_0_woDiffuse_wNoiseresampling_MockSim_Distance5.5kpc_Nmax_500.pdf}\hfill
    \includegraphics[width=0.49\columnwidth]{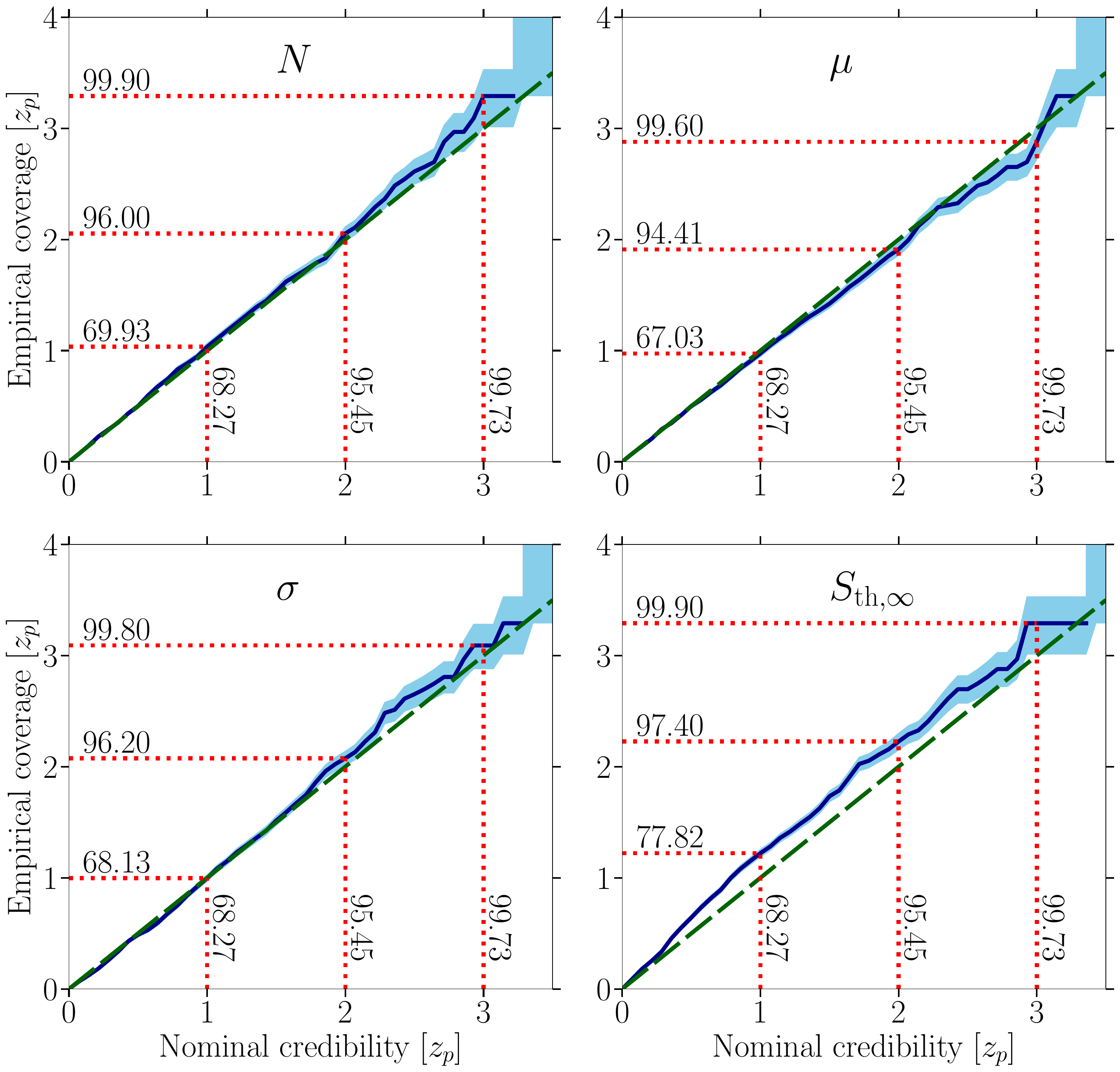}
    \caption{(\emph{Left}:) Same as Figure~\ref{fig:singleGC-pflux0p0_wodiff} using a wider prior on $N$ extending up to 3000 MSPs. (\emph{Right}:) Coverage plot for the scenario considered in the left panel. \label{fig:wide-priors-baseline}}
\end{figure*}

In this appendix section, we show the results of enlarging the prior range of the \sbi~analysis of our baseline mock dataset. We increase the total numbers of sources from $N = 500$ to $N = 3000$ while keeping the priors of all other parameters identical to the default setup shown in Table~\ref{tab:swyft_priors}. The posteriors and the coverage for this case are displayed in Figure~\ref{fig:wide-priors-baseline}.

\section{Updating the distance to Terzan 5}
\label{app:ter5-new-distance}

\begin{figure*}
    \centering
    \includegraphics[width=0.49\columnwidth]{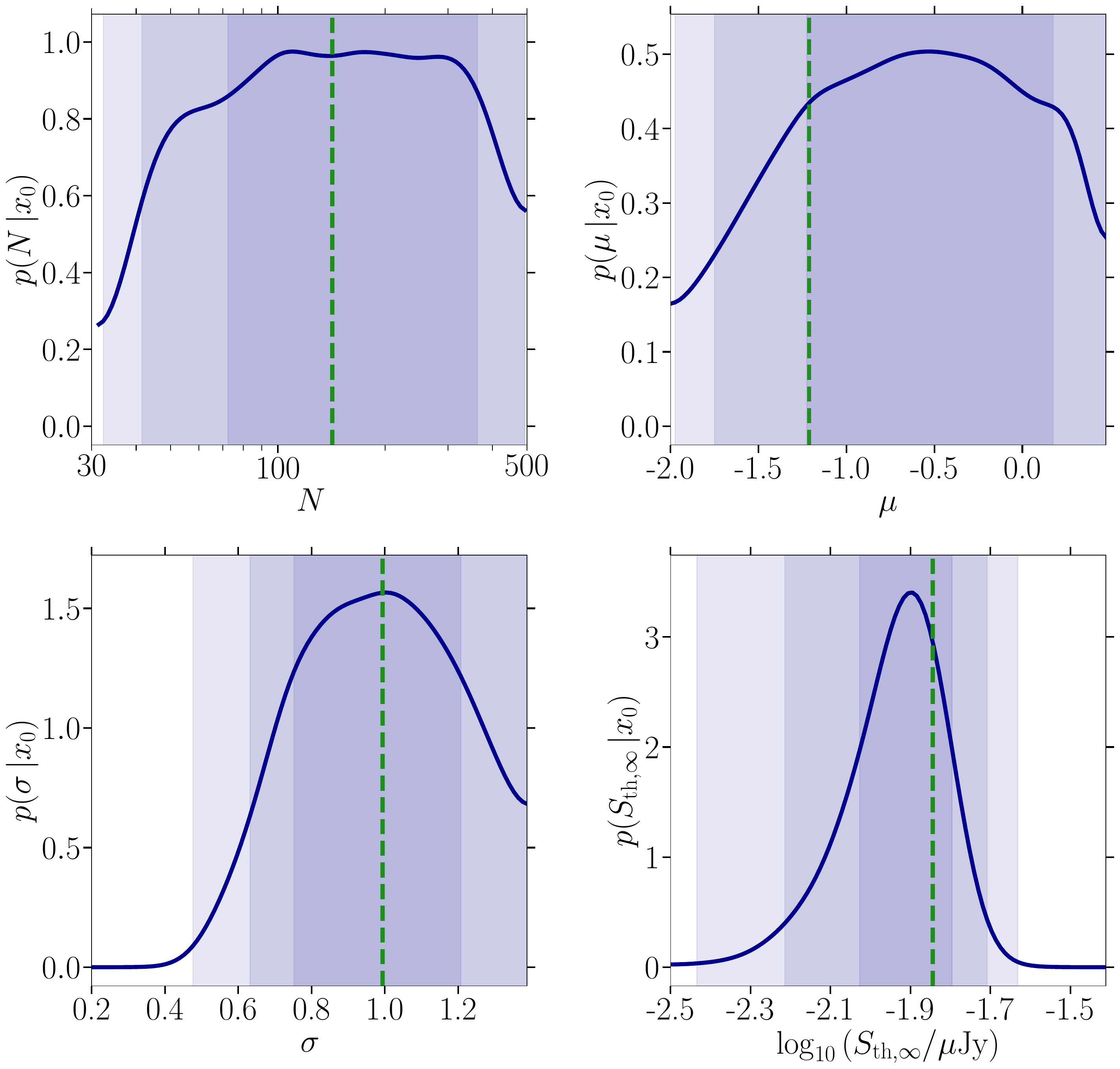}\hfill
    \includegraphics[width=0.49\columnwidth]{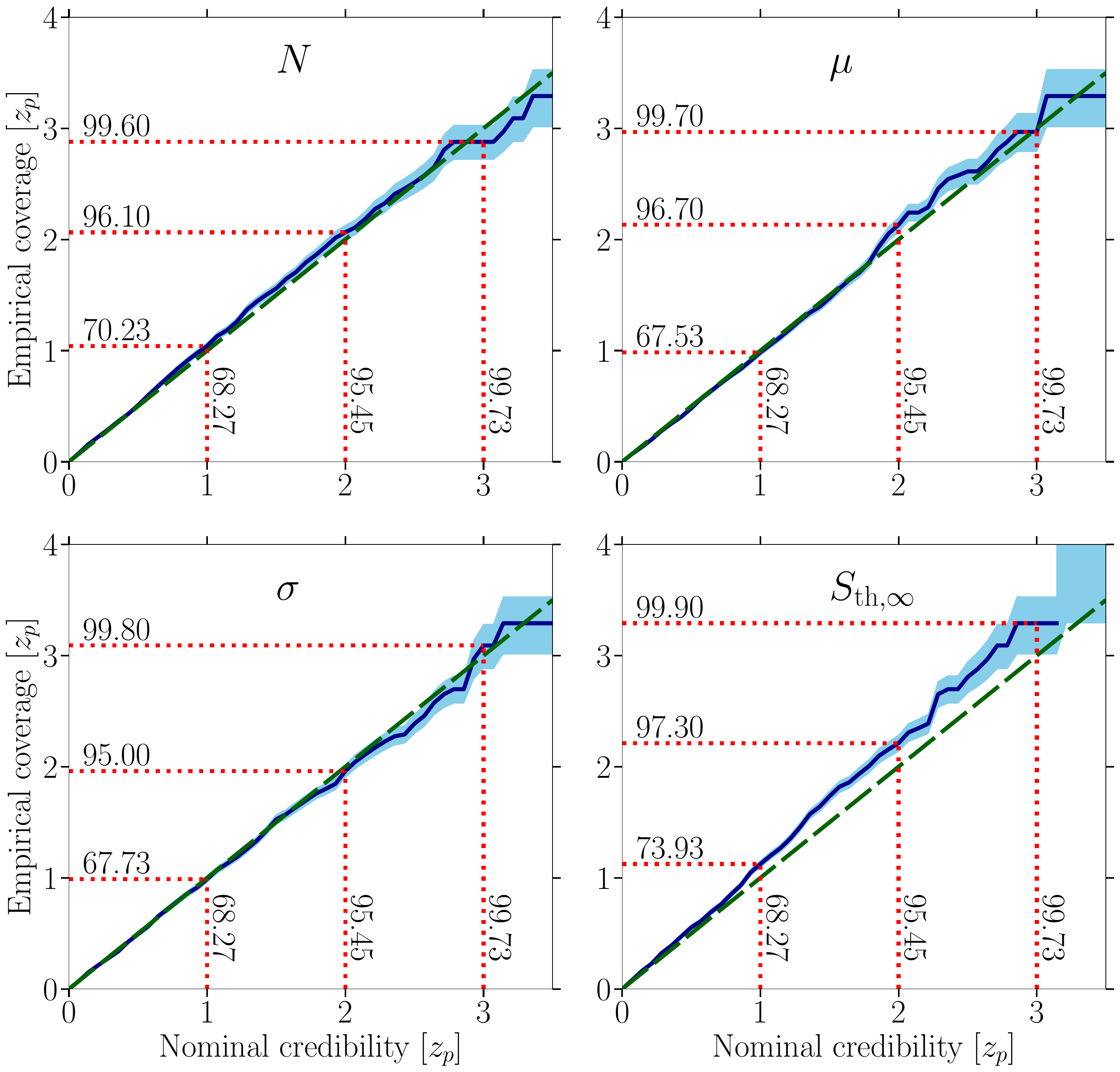}
    \caption{(\emph{Left}:) Same as Figure~\ref{fig:swyft-terzan5-results} using the updated distance to Ter 5 of about 6.62 kpc and the GBT sample of $N_{\mathrm{detected}} = 31$. (\emph{Right}:) Coverage plot for the scenario considered in the left panel. \label{fig:ter5-new-distance}}
\end{figure*}

In this appendix section, we show the results of changing the distance to Ter 5 from 5.5 kpc to 6.62 kpc as inferred from datasets of the \textit{Gaia} mission in \cite{2021MNRAS.505.5957B}. We inspect the case of $n = 32$, that is, the Ter 5 GBT sample where we show the posteriors and the coverage for this case in Figure~\ref{fig:ter5-new-distance}.

\section{Multiple clusters}
\label{app:multiple_gc}

In Sections \ref{sec:mock_data_bayes} and \ref{sec:bayes_framework}, we introduced several quantities related either to pulsars or to their host cluster, Terzan 5. In order to extend our analysis to several cluster simultaneously, one can use exponents to indicate cluster-dependence. As an example, equation \ref{eq:logl_to_logs} becomes:
\begin{equation}
    \log_{10}(S_i^j) = \log_{10}(L_i^j) - 2 \log_{10}(d^j).
\end{equation}
If the luminosity function is assumed to be the same for all clusters, and therefore, that $\mu$ and $\sigma$ are unique, Equation \ref{eq:logl_pdf} remains valid, and each new cluster only adds 3 new parameters to the model: $N^j$, $S_\mathrm{th}^j$ and $d^j$. Otherwise, $f$, $\mu$ and $\sigma$ simply become $f^j$, $\mu^j$ and $\sigma^j$, and each new cluster only adds 5 new parameter to the model. Equation \ref{eq:bayes} can be used in both cases with $\mathcal{L}(D|\theta, M)$ computed as the product of the $\mathcal{L}^j(D^j|\bm{\theta}^j, M)$.

\end{document}